\documentclass[11pt]{article}
\RequirePackage{fullpage}

\RequirePackage{enumerate}
\RequirePackage[latin1]{inputenc}
\RequirePackage{amsmath}
\RequirePackage{amssymb}

\usepackage{hyperref}
\hypersetup{
  colorlinks, linkcolor=blue
}

\usepackage{algorithmicx}
\usepackage{algorithm}
\usepackage[noend]{algpseudocode}

\algdef{SE}[DOWHILE]{Do}{doWhile}{\algorithmicdo}[1]{\algorithmicwhile\ #1}

\usepackage{amsmath,amsfonts}
\usepackage[colorinlistoftodos]{todonotes}
\usepackage{epsfig}
\newcommand{\NP}{NP}

\newcommand{\qed}{\hfill$\diamond$}
\newcommand{\pf}{{\textbf {Proof:} }}
\newtheorem{theorem}{Theorem}[section]

\newtheorem{lemma}[theorem]{Lemma}

\newtheorem{claim}[theorem]{Claim}
\newtheorem{obs}[theorem]{Observation}

\newtheorem{dfn}[theorem]{Definition}

\newcommand{\beq}{\begin{equation}}
\newcommand{\eeq}{\end{equation}}

\newcommand{\hidetext}[1]{}

\begin{document}

\title{Digraph homomorphism problems and weak near unanimity polymorphisms}

\author{
  Tom\'{a}s Feder\thanks{268 Waverley Street, Palo Alto, CA 94301, United States, tomas@theory.stanford.edu}   
      \and 
       Jeff Kinne \thanks{Indiana State University, IN, USA, jkinne@cs.indstate.edu, Supported by NSF grant 1751765} 
      \and 
        Ashwin Murali \thanks{Indiana State University, IN, USA, amurali1@sycamores.indstate.edu, supported by NSF 1751765}
        \and
   Arash Rafiey \thanks{Indiana State University, IN, USA arash.rafiey@indstate.edu and Simon Fraser University, BC, Canada, arashr@sfu.ca, Supported by NSF grant 1751765 }}


\date{}
\maketitle
\begin{abstract}
 We consider the problem of finding a homomorphism from an input digraph $G$ to a fixed digraph $H$. We show that if $H$
admits a weak near unanimity polymorphism $\phi$ then deciding whether $G$ admits a homomorphism to $H$ (HOM($H$))
is polynomial time solvable. This gives a  proof of the dichotomy conjecture (now dichotomy theorem) by Feder
and Vardi \cite{FV93}. Our approach is combinatorial, and it is simpler than the two algorithms found by Bulatov \cite{bulatovCSP} and Zhuk \cite{zhukCSP} in 2017. We have implemented our algorithm and show some experimental results. We use our algorithm together with the recent result \cite{arash-maltsev} for recognition of Maltsev polymorphisms and decide in polynomial time if a given relational structure $\mathcal{R}$ admits a weak near unanimity polymorphism.




\end{abstract}


\section{Introduction} \label{apsec:background}





For a digraph $G$, let $V(G)$ denote the vertex set of $G$ and let $A(G)$ denote
the arcs (aka edges) of $G$.  An arc $(u,v)$ is often written as simply $uv$ to shorten expressions. Let $|G|$ denote the number of vertices in $G$. 

A {\em homomorphism} of a digraph $G$ to a digraph $H$ is a mapping $g$ of the vertex set of $G$ to the vertex set of $H$
so that for every arc $uv$ of $G$ the image $g(u)g(v)$ is an arc of $H$. A natural decision problem is whether for given
digraphs $G$ and $H$ there is a homomorphism of $G$ to $H$.  If we view (undirected) graphs as digraphs in which each edge
is replaced by the two opposite directed arcs, we may apply the definition to graphs as well. An easy reduction from the
$k$-coloring problem shows that this decision problem is $\NP$-hard: a graph $G$ admits a $3$-coloring if and only if there is
a homomorphism from $G$ to $K_3$, the complete graph on $3$ vertices. As a homomorphism is easily verified if the mapping
is given, the homomorphism problem is contained in $\NP$ and is thus $\NP$-complete.

The following version of the problem has attracted much recent attention. For a fixed digraph $H$ the problem $HOM(H)$ asks
if a given input digraph $G$ admits a homomorphism to $H$. Note that while the $3$-coloring reduction shows $HOM(K_3)$ is NP-complete,
$HOM(H)$ can be easy (in $P$) for some graphs $H$: for instance if $H$ contains a vertex with a self-loop, then every graph $G$ admits a homomorphism
to $H$. Less trivially, for $H = K_2$ (or more generally, for any bipartite graph $H$), there is a homomorphism from $G$ to $K_2$ if
and only if $G$ is bipartite. A very natural goal is to identify precisely for which digraphs $H$ the problem $HOM(H)$ is easy. In the
special case of graphs the classification has turned out
to be this: if $H$ contains a vertex with a self-loop or is bipartite, then $HOM(H)$ is in $P$, otherwise it is $NP$-complete \cite{hell-nesetril}
(see \cite{sigge} for shorter proofs). This classification result implies a {\em dichotomy} of possibilities for the problems $HOM(H)$ when $H$
is a graph, each problem being $NP$-complete or in $P$. However, the dichotomy of $HOM(H)$ remained open for
general digraphs $H$ for a long time. It was observed by Feder and Vardi \cite{FV93} that this problem is equivalent to the dichotomy of a much
larger class of problems in $NP$, in which  $\mathcal{H}$ is a fixed finite relational structure. These problems can be viewed as {\em constraint
satisfaction problems} with a fixed template $\mathcal{H}$ \cite{FV93}, written as CSP($\mathcal{H}$). 

The CSP($\mathcal{H}$) involves deciding, given a set
of variables and a set of constraints on the variables, whether or not there is an assignment (form the element of $H$) to
the variables satisfying all of the constraints. 

This problem can be formulated in terms of homomorphims as follows. Given a pair $(\mathcal{G}, \mathcal{H})$ of \emph{relational structures}, decide whether or not there is
a homomorphism from the first structure to the second structure. 


3SAT is a prototypical instance of CSP, where each variable takes values of {\em true} or {\em false} (a domain
size of two) and the clauses are the constraints.  Digraph homomorphism problems can also easily be converted into CSPs: the variables $V$
are the vertices of $G$, each must be assigned a vertex in $H$ (meaning a domain size of $|V(H)|$), and the constraints encode that each
arc of $G$ must be mapped to an arc in $H$.

Feder and Vardi argued in \cite{FV93} that
in a well defined sense the class of problems $CSP(H)$ would be the largest subclass of $NP$ in which a dichotomy holds. A
fundamental result of Ladner \cite{L75} asserts that if $P \ne NP$ then there exist $NP$-intermediate problems (problems neither
in $P$ nor $NP$-complete), which implies that there is no such dichotomy theorem for the class of {\em all} $NP$ problems.  Non-trivial
and natural sub-classes which do have dichotomy theorems are of great interest. Feder and Vardi made the following {\em Dichotomy
Conjecture}: every problem $CSP(H)$ is $NP$-complete or is in $P$. This problem has animated much research in theoretical computer
science. For instance the conjecture has been verified when $H$ is a conservative relational structure \cite{bulatov}, or a digraph with all in-degrees
and all-out-degrees at least one \cite{barto}. Numerous special cases of this conjecture have been verified \cite{ABISV09,BHM88,BH90,B06, CVK10, D00, F01, F06, FMS04, LZ03, schaefer}.

Bulatov gave an algebraic proof for the conjecture in 2017 \cite{bulatovCSP} and later Zhuk \cite{zhukCSP} also announced another algebraic proof of the conjecture.



It should be remarked that constraint satisfaction problems encompass many well known computational problems,
in scheduling, planning, database, artificial intelligence, and constitute an important area of applications, in addition to
their interest in theoretical computer
science \cite{CKS01,D92,K92,V00}.

While the paper of Feder and Vardi \cite{FV93} did identify some likely candidates for the boundary between easy and hard 
$CSP$-s, it was the development of algebraic techniques by Jeavons \cite{jeavons}
that lead to the first proposed classification \cite{bjk}.
The algebraic approach depends on the observation that the complexity of $CSP(H)$ only depends on certain symmetries of $H$, the
so-called {\em polymorphisms} of $H$. For a digraph $H$ a polymorphism $\phi$ of arity $k$ on $H$ is a homomorphism from $H^k$ to $H$. Here 
$H^k$ is a digraph with vertex set $\{(a_1,a_2,\dots,a_k) | a_1,a_2,\dots,a_k \in V(H)\}$ and arc set
$\{(a_1,a_2,\dots,a_k)(b_1,b_2,\dots,b_k) \ \ | \ \  \ \ a_ib_i \in A(H) \textnormal{ for all } 1 \le i \le k\}$. For a polymorphism $\phi$,  $\phi(a_1,a_2,\dots,a_k)\phi(b_1,b_2,\dots,b_k)$ is an arc of $H$ whenever
$(a_1,a_2,\dots,a_k)(b_1,b_2,\dots,b_k)$ is an arc of $H^k$. 

Over time, one concrete classification has emerged as the likely
candidate for the dichotomy. It is expressible in many equivalent ways, including the first one proposed in \cite{bjk}.
 There were thus a number of equivalent conditions on $H$ that were postulated to describe which problems
$CSP(H)$ are in $P$. For each, it was shown that if the condition is not satisfied then the problem $CSP(H)$ is $NP$-complete
(see also the survey \cite{ccc}).
One such condition is the existence of a weak near unanimity polymorphism (Maroti and McKenzie \cite{maroti}). 
A polymorphism $\phi$ of $H$ of arity $k$ is a {\em $k$-near unanimity function}
($k$-NU) on $H$, if $\phi(a,a,\dots,a)=a$ for every $a \in V(H)$, and
$\phi(a,a,\dots,a,b)=\phi(a,a,\dots,b,a)=\dots=\phi(b,a,\dots,a)=a$ for every
$a, b \in V(H)$. If we only have $\phi(a,a,\dots,a)=a$ for every $a \in V(H)$ and
$\phi(a,a,\dots,a,b)=\phi(a,a,\dots,b,a)=\dots=\phi(b,a,\dots,a)$ [not necessarily $a$] for every $a,b \in V(H)$, then $\phi$ is a {\em weak $k$-near unanimity function} (weak $k$-NU). A polymorphism $\phi$ on $H$ is \emph{Siggers} if $\phi(a,r,e,a)=\phi(r,a,r,e)$ for every $a,r,e \in H$ \cite{Siggers}. It has been shown by Siggers \cite{Siggers} that if digraph $H$ admits a Sigger polymorphism it  also admits a weak $k$-NU for some $k  \ge 3$.

Given the $NP$-completeness proofs that are known, the proof of the Dichotomy
Conjecture reduces to the claim that a relational structure $\mathcal{H}$ which admits a weak near unanimity polymorphism has a polynomial time
algorithm for $CSP(\mathcal{H})$. As mentioned earlier, Feder and Vardi have shown that is suffices to prove this for $HOM(H)$ when $H$ is a
digraph. This is the main result of our paper. 

Note that the real difficulty in the proof of the \emph{graph} dichotomy theorem in \cite{hell-nesetril} lies in proving the $NP$-completeness. By
contrast, in the \emph{digraph} dichotomy theorem proved here it is the polynomial-time algorithm that has proven more difficult.

While the main approach in attacking the conjecture has mostly been to use the highly developed techniques from logic and algebra, and
to obtain an algebraic proof, we go in the opposite direction and develop a combinatorial algorithm. Our main result is the following.
\begin{theorem} \label{main-theorem}
Let $H$ be a digraph that admits a weak near unanimity function. Then $HOM(H)$ is in $P$.  
Deciding whether an input digraph $G$ admits a homomorphism to $H$ can be done in time $\mathcal{O}(|G|^ {4}|H|^{k+4})$. 
\end{theorem}


\paragraph{Very High Level View}
We start with a general digraph $H$ and a weak $k$-NU $\phi$ of $H$. 
We turn the problem $HOM(H)$ into a related problem of seeking a homomorphism with lists of allowed images. The {\em list homomorphism
problem} for a fixed digraph $H$, denoted $LHOM(H)$, has as input a digraph $G$, and for each vertex $x$ of $G$ an associated list (set)
of vertices $L(x) \subseteq V(H)$, and asks whether there is a homomorphism $g$ of $G$ to $H$ such that for each $x\in V(G)$, the image of $x$; 
$g(x)$, is in $L(x)$.
Such a homomorphism is called a {\em list homomorphism} of $G$ to $H$ with respect to the lists $L$. 
List homomorphism problems have been extensively studied, and are known to have nice dichotomies \cite{FH98,FHH03,soda11}. However, we can not use the algorithms for finding list homomorphism from $G$ to $H$, because in the $HOM(H)$ problem, for every vertex $x$ of $G$, $L(x)=V(H)$. 



\paragraph{Preprocessing} 
One of the common ingredients in $CSP$ algorithms is the use of consistency checks to
reduce the set of possible values for each variable (see, for example the algorithm
outlined in \cite{pavol-book} for $CSP(H)$ when $H$ admits a 
near unanimity function).
Our algorithm includes such a consistency check (also known as 
(2,3)-consistency check \cite{FV93})  as a first step which we call \emph{PreProcessing}. PreProcessing procedure begins by performing arc and pair consistency check on the list of vertices
in the input digraph $G$. 
For each pair $(x,y)$ of $V(G) \times V(G)$ we consider a list of possible pairs $(a,b)$,
$a \in L(x)$ (the list in $H$ associated with $x \in V(G)$) and $b \in L(y)$.
Note that if $xy$ is an arc of $G$ and $ab$ is not an arc of $H$
then we remove $(a,b)$ from the list of $(x,y)$.
Moreover, if $(a,b) \in L(x,y)$ and there exists $z$ such that there
is no $c$ for which $(a,c) \in L(x,z)$ and $(c,b) \in L(z,y)$
 then we remove $(a,b)$ from the list of $(x,y)$.
 We continue this process until no list can be modified.
 If there are empty lists then clearly there is no list homomorphism from $G$ to $H$. 

\paragraph{After PreProcessing} The main structure of the algorithm is to perform {\em pairwise elimination}, which focuses on two vertices $a, b$ of $H$ that occur together in some list
$L(x), x \in V(G)$, and finds a way to eliminate $a$ or $b$ from $L(x)$ without changing a feasible problem into an unfeasible one. In other
words if there was a list homomorphism with respect to the old lists $L$, there will still be one with respect to the updated lists $L$. This process
continues until either a list becomes empty, certifying that there is no homomorphism with respect to $L$ (and hence no homomorphism at all),
or until all lists become singletons specifying a concrete homomorphism of $G$ to $H$ or we reach an instance that has much simpler structure and can be solved by the existing CSP algorithms. This method has been successfully used in other papers \cite{soda14,soda11,mincost-dichotomy}.

In this paper, the choice of which $a$ or $b$ is eliminated, and how,
is governed by the given weak near unanimity
polymorphism $\phi$. The heart of the algorithm is a delicate procedure for updating the lists $L(x)$  in such a way that (i) feasibility is maintained, and the polymorphism $f$ keep its initial property (which is key to maintaining feasibility).

\paragraph{Meta-Question} An interesting question arising from  the study of CSP Dichotomy theorem is known as the \emph{meta-question}. Given a relational structure $\mathcal{H}$, decide whether or not $\mathcal{H}$ admits a polymorphism from a class--for various classes of polymorphims. For many cases hardness results are known. Semmilattice, majority, Maltsev, NU, and weak NU, are among the popular polymorphisms when it comes to study of CSP. Having one or more of these polymorphisms on relation $\mathcal{H}$, would make the  CSP($\mathcal{H}$) (or variation) instance tractable. Therefore, knowing structural characterization and polynomial time recognition for these polymorphisms would help in designing efficient algorithms for solving CSP. A binary polymorphism $f$ on $\mathcal{H}$ is called \emph{semilattice} if $f(a,b)=f(b,a)$, and $f(a,a)=a$, $f(a,f(a,c))=f(f(a,b),c)$ for every $a,b,c \in V(\mathcal{H})$. A ternary polymorphism $g$ on $\mathcal{H}$ is \emph{majority} if $g(a,a,a)=g(a,a,b)=g(a,b,a)=g(b,a,a)=a$.  A polymorphism $h$ of arity three on $\mathcal{H}$ is called \emph{Maltsev} if for every $a,b \in V(\mathcal{H})$, 
 $h(a,a,a)=h(a, b, b) = h(b,b, a) = a$.  

It was shown in \cite{benoit} that deciding if a relational structure admits any of the following polymorphism is NP-complete;  a semilattice polymorphism, a conservative semilattice polymorphism,  a commutative, associative polymorphism (that is, a commutative semigroup polymorphism). However, when $\mathcal{H}$ is a single binary relation(digraph) then deciding whether $\mathcal{H}$ admits a conservative semmilattice is polynomial time solvable \cite{bi-arc}. Relational structure and digraphs with majority/ near unanimity polymorphism have studied in  \cite{nuf1,benoit,nuf2,soda11,kazda,maroti}. 

One remaining open question is whether the existence of a weak NU polymorphism for a given relational structure can be decided in polynomial time \cite{bkw}. We transform this problem into a graph list homomorphism problem from an input graph $G$ to a target graph $H$, in which $G \times H^4$ admits a Siggers polymorphism (i.e. 
a weak NU polymorphism of arity $k$ for some $k \ge 3$) with respect to the lists. To solve this problem, we need some modules of our algorithm for finding a homomorphism from $G$ to $H$ when $H$ admits a weak NU. Moreover, we also need an algorithm for solving the list homomorphism problem from an input graph $G$ to a target graph $H$ where 
$G \times H^3$ admits a Maltsve polymorphism with respect to the lists; such an algorithm would only assume the existence of a Maltsev polymorphism without knowing the actual values. The later algorithm has been designed in \cite{arash-maltsev}. Deciding whether a relational structure admits a Malstve polymorphims has been an open problem \cite{bkw,Siggers}.  We prove the following theorem. 

\begin{theorem}\label{weak-nu-recognition} 
Let $\mathcal{R}$ be a relational structure. Then the problem of deciding whether $\mathcal{R}$ admits a weak NU polymorphism is polynomial time solvable. 
\end{theorem}

\subsection{ Addressing the issue with the 2017 manuscript}


Our algorithm (in manuscript 2017) makes a decision based on the output of a test $T_{x,b,a}$ on a smaller instance of the digraph homomorphism problem. Here $a,b$ are possible images for $x \in V(G)$, of a homomorphism from $G$ to $H$. 

The algorithm assumes that the test $T_{x,b,a}$ outputs``yes'' and based on the correctness of the test, $a$ is removed from further consideration for $x$. The test $T_{x,b,a}$ uses the properties of the weak NU polymorphism $\phi$. However, it is conceivable that the test $T_{x,b,a}$ fails, and this means we should not remove $a$ from the list of possible images of $x$.  We had incorrectly claimed in the manuscript that the properties of $\phi$ and pre-tests in the algorithm guarantee the test always passes. But we can construct such an example where the test must fail in the algorithm as follows. 
Let $H$ be a digraph with two weakly connected components $H_1,H_2$. The weak NU polymorphism $\phi$ could be of arity three and such that for every $a \in H_1$ and every $b \in H_2$,  $\phi(a,b,b)=\phi(b,b,a)=\phi(b,a,b)=c$ for some $c \in H_2$. Suppose there exists a homomorphism from $G$ (weakly connected) to $H$ that maps $x$ to $a$ and hence the entire graph $G$ must be mapped to $H_1$. Moreover, suppose there is no homomorphism from $G$ to $H_2$. The algorithm does consider the test $T_{x,b,a}$ eventually for such $G$ and $H$.  According to the test $T_{x,b,a}$, we remove $a$ from further consideration for $x$ which leads us to remove the possible homomorphism from $G$ to $H$.  

Note that one can assume $H$ is weakly connected as follows. Suppose $H_1, H_2$ are balanced digraphs with $\ell$ levels (we can partitioned the vertices of $H_i$, $i=1,2$, into $\ell$ parts where all the arcs of $H_i$ go from a vertex in some part $j$ to part $j+1$). An extra vertex $a'$ can be added and connected to all the vertices of $H_1, H_2$ on the lowest level. This way we obtain the weakly connected digraph $H = H_1 \cup H_2 \cup \{a'\}$ with $\ell+1$ levels. We may assume $G$ is also balanced and has $\ell$ levels. 
An extra vertex $x'$ can be added to $G$ with arcs to every vertex of $G$ on the lowest level. Now $G'= G \cup \{x'\}$ is also a balanced digraph with $\ell+1$ levels. Note that in any homomorphism from $G'$ to $H$, $x'$ must be mapped to $a'$ and any other vertex of $G'$ must map to $H-\{x\}$. 

We note that Ross Willard has posted a concrete counter-example (and further discussion of his example) for which $H$ contains 197 vertices in $H$. The example inspired by instances of the CSP, so-called Semi-lattice block Maltsev. In Subsection \ref{example} and Section \ref{experiment}, we discuss these kinds of examples in detail and show how our new algorithm handles these examples.

\section{Necessary Definitions}

An {\em oriented walk (path)} is obtained from a walk (path) by orienting each of its edges. The {\em net-length} of a walk $W$, is
the number of forward arcs minus the number of backward arcs following $W$ from the beginning to the end.
An {\em oriented cycle} is obtained from a cycle by orienting each of its edges. We say two oriented walks $X,Y$ are congruent if 
they follow the same patterns of forward and backward arcs. 

For $k$ digraphs $G_1,G_2,\dots,G_k$, let $G_1 \times G_2 \times \dots \times G_k$ be the digraph with vertex set $\{(x_1,x_2,\dots,x_k) | x_i \in V(G_i), 1 \le i \le k\}$ and arc set $\{(x_1,x_2,\dots,x_k)(x'_1,x'_2,\dots,x'_k) | x_ix'_i \in A(G_i), 1 \le i \le k\}$. Let $H^k = H \times H \times \dots H$, $k$ times.  

Given digraphs $G$ and $H$, and $L: G \rightarrow 2^H$,  let 
$G \times_L H^k$ be the induced sub-digraph of $G \times H^k$ with the vertices $(y;a_1,a_2,\dots,a_k)$ 
where $a_1,a_2,\dots,a_k \in L(y)$.

\begin{dfn}[Homomorphism consistent with Lists]
\label{dfn:homomorphism-main}
  Let $G$ and $H$ be digraphs and list function $L : V(G) \rightarrow 2^{H}$, i.e. list of $x \in V(G)$, $L(x) \subseteq V(H)$. Let $k>1$ be  an integer.

  A function $f: G \times_L H^k \rightarrow H$ is a \emph{list homomorphism with respect to lists $L$ } if the following
  hold.
  \begin{itemize}

  \item \emph{List property :}
   for every $(x;a_1,a_2,\dots,a_k) \in G \times_L H^k$, $f(x;a_1,a_2,\dots,a_k) \in L(x)$

  \item \emph{Adjacency property:}
     if $(x;a_1, ..., a_k)(y; b_1, ..., b_k)$ is an arc of $G \times_L H^k$ then \\
    $f(x; a_1, ..., a_k)f(y; b_1, ..., b_k)$ is an arc of $H$.

  \end{itemize}
\end{dfn}

In addition if $f$ has the following property then we say $f$ has the {\em weak $k$-NU property}. 

\begin{itemize}
 \item  for every $x\in V(G), \{a, b\} \subseteq L(x)$, we have
    $f(x; a, b, b, ..., b) = f(x; b, a, b, ..., b) = ... = f(x; b, b, b, ... a)$. 
    \item for every $x \in V(G)$, $a \in L(x)$, we have $f(x;a,a,\dots,a)=a$.
\end{itemize}

We note that this definition is tailored to our purposes and in particular differs from the standard definition of weak $k$-NU as follows. $f$ is based on two
  digraphs $G$ and $H$ rather than just $H$ (we think of this as starting with a
  traditional weak $k$-NU on $H$ and then allowing it to vary somewhat for each
  $x \in V(G)$).

\paragraph{Notation} 
For simplicity let $(b^k,a)= (b,b,\dots,b,a)$ be a $k$-tuple of all $b$'s but with an $a$ in the $k^{th}$ coordinate. Let $(x;b^k,a)$ be a $(k+1)$-tuple of $x$, $(k-1)$ $b$'s and  $a$ in the $(k+1)^{th}$ coordinate.

\begin{dfn}[$f$-closure of a list]\label{rec-closure} We say a set $S \subseteq L(y)$ is closed under $f$ if for every $k$-tuple $(a'_1,a'_2,\dots,a'_k) \in S^k$ we have  $f(y;a'_1,a'_2,\dots,a'_k) \in S$. 
For set $S \subseteq L(y)$, let $\widehat{f}_{y,S} \subseteq L(y)$ be a minimal set that includes all the elements of $S$ and it is closed under $f$. 
\end{dfn}

Let $X : x_1,x_2,\dots,x_n$ be an oriented path in $G$. Let $X[x_i,x_j]$, $1 \le i \le j \le n$, denote the induced sub-path of $X$ from $x_i$ to $x_j$. Let $L(X)$ denote the vertices of $H$ that lie in the list of the vertices of $X$.   

\begin{dfn}[induced  bi-clique]
We say two vertices $x,y$ induced a bi-clique if there exist vertices $a_1,a_2,\dots,a_r \in L(x)$, $r>1$ and $b_1,b_2,\dots,b_s \in L(y)$ such that $(a_i,b_j) \in L(x,y)$ for every $1 \le i \le r$ and $1 \le j \le s$. 


\end{dfn}

   \begin{figure}
   \begin{center}
 \includegraphics[scale=0.60]{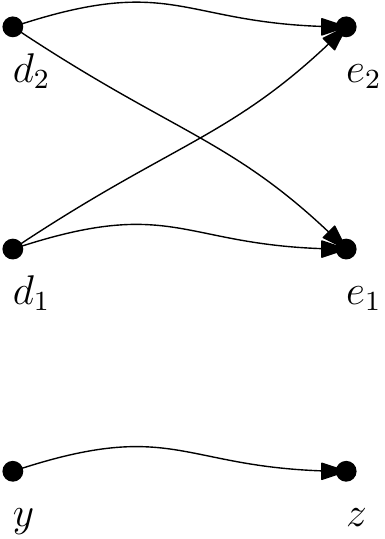}
   \end{center}
   \caption{  An example of a Bi-clique}
 \label{fig:Bi-clique-example}
  \end{figure}

Let $a_1,a_2 \in L(x)$ and suppose there exist $b_1,b_2 \in L(y)$ such that $(a_1,b_1),(a_1,b_2),(a_2,b_1),  (a_2,b_2) \in L(x,y)$. Then it follows from the property of $f$, that $\widehat{f}_{x,\{a_1,a_2\}}$
and $\widehat{f}_{y,\{b_1,b_2\}}$ induce a bi-clique on $x,y$. 

\begin{dfn}[weakly connected component in lists $L$ ] 

By \emph{connected component} of $G \times_L H$ we mean a \emph{weakly connected component} $C$ of digraph $G \times_L H$(i.e. a connected component of $G \times_L H$ when we ignore the direction of the arcs) which is closed under $(2,3)$-consistency. That means, for every $(x,a),(y,b) \in C$, and every $z \in V(G)$ there is some $c \in L(z)$ such that $(a,c) \in L(x,z)$, $(b,c) \in L(y,z)$.  
 
\end{dfn}

\begin{obs} \label{obs1}
If there exists a homomorphism $g : G \rightarrow H$ then all the vertices $(y,g(y))$, $y \in V(G)$ 
belong to the same connected component of $G \times_L H$. 
\end{obs}

\begin{dfn}
For $a,b \in L(x)$ we say $(b,a)$ is a non-minority pair if $f(x;b^k,a) \ne a$. Otherwise, we say $(b,a)$ is a minority pair. 
\end{dfn}

\begin{dfn}
For $x \in V(G)$, $a \in L(x)$, let  $L_{x,a}$ be the subset of lists $L$ that are consistent with $x$ and $a$. In other words,  
for every $y \in V(G)$, $L_{x,a}(y)= \{ b \in L(y)  | (a,b) \in L(x,y) \}$. Note that by definition $L_{x,a}(x)=\{a\}$. In general  for $x_1,x_2,\dots,x_t \in V(G)$, let $L_{x_1,a_1,x_2,a_2,\dots,x_t,a_t}$ be the subset of lists $L$ that are consistent 
with all the  $a_i \in L(x_i)$'s, $1 \le i \le t$.  In other words, for every $y \in V(G)$, 
$L_{x_1,a_1,x_2,a_2,\dots,x_t,a_t}(y) = \{ b \in L(y) | (a_i,b) \in L(x_i,y), i=1,2,\dots,t \}$.   

\end{dfn}

\section{Algorithm} \label{sec:procedure}

The main algorithm starts with applying the Preprocessing procedure on the instance $G,H,L,\phi$, where $\phi$ is a weak NU polymorphism of arity $k$ on $H$. If we encounter some empty (pair) lists then there is no homomorphism from $G$ to $H$, and the output is no. Otherwise, it defines the weak NU list homomomorphism $f: G \times_L H^k \rightarrow H$, by setting $f(x;a_1,a_2,\dots,a_k)=\phi(a_1,a_2,\dots,a_k)
$ for every $x \in G$ and every $a_1,a_2,\dots,a_k \in L(x)$. Then it proceeds with the \Call{Not-Minority}{} algorithm (Algorithm \ref{alg-not-minority}). Inside the \Call{Not-Minority}{} algorithm we look for the special cases, the so-called minority cases which are turned into a Maltsev instances, and can be handled using the existing Maltsev algorithms.

\begin{algorithm}
\caption{The main algorithm for solving the digraph homomorphism problem}
  \label{rec-alg-main}
  \begin{algorithmic}[1]
     \State \textbf{Input:} Digraphs $G,H$, and, a weak NU homomorphism $\phi :H^k \rightarrow H$ 
    
    \Function{DigraphHom}{$G, H, \phi$}
   
    \State \textbf{for all} $x\in V(G)$, let $L(x) = V(H)$
  
 \If { \Call{PreProcessing}{$G, H, L$} is false} return "no homomorphism"
 \EndIf
     
       \State \textbf{for all} $x \in V(G)$ and $a_1, ..., a_k \in L(x)$, let $f(x; a_1, ..., a_k) = \phi(a_1, ..., a_k)$
     
     \State $g=$\Call{Not-Minority}{$G,H,L,f$}

   \If { $g$ is not empty  } {\textbf {return}} true 
    
   \Else { } {\textbf {return}} false 
   
   \EndIf
  
    \EndFunction

  \end{algorithmic}

\end{algorithm}

 \begin{algorithm}[ht]
 \begin{algorithmic}[1] 
 
 \State \textbf{Input:} Digraphs $G,H$, lists $L$ and, a weak NU homomorphism $f : G \times_L H^k \rightarrow H$ which is minority 
      \Function{RemoveMinority} {$G,H,L,f$ } \label{RemoveMinority}

      \ForAll { $x \in V(G)$, and $a,b,c \in L(x)$   }
              
   \State Set $h(x;a,b,c)=f(x;a,b,b,\dots,b,c)$
  \EndFor

   \State $g$=\Call{Maltsev-Algorithm}{$G,H,L,h$}
   \State \textbf {return} $g$
 \EndFunction

 \end{algorithmic}

 \end{algorithm}

\paragraph{Minority Instances} 
Inside function \Call{Not-Minority}{} we first check whether the instance is Maltsev or {\em Minority} instance -- 
in which we have a homomorphism $f$ consistent with $L$ such that
for every $a,b \in L(x)$, $(b,a)$ is a minority pair, i.e. 
$f(x;b^k , a)=a$, and in particular when $a=b$ we have $f(x;a,a,\dots,a)=a$ (idempotent property). 

In our setting a homomorphism $h : G \times_L H^3 \rightarrow H$ is called Maltsev list homomorphism if $h(x;a,a,b)=h(x;b,a,a)=b$ for every $a,b \in L(x)$, $x \in V(G)$.  

Let $G,H,L,f$ be an input to our algorithm, and suppose all the pairs are minority pairs.  
We define a homomorphism $h :G \times_L H^3 \rightarrow H$ (consistent with $L$) by setting $h(x;a,b,c)=f(x;a,b,b,\dots,b,c)$ for
$a,b,c \in L(x)$. Note that since $f$ has the minority property for all $x \in V(G)$, $a, b\in L(x)$, $h$ is a Maltsev homomorphism consistent with the lists $L$. This is because when $b=c$ then $h(x;a,b,b)=f(x;a,b,\dots,b)=a$, and when $a=b$, $h(x;a,a,c)=f(x;b,b,\dots,b,c)=c$, for every $a,c$, and hence, $h(x;b,b,a)=a$.  

The Maltsev/Minority instances can be solved using the algorithm in \cite{BD06}. Although the algorithm in \cite{BD06} assumes there is a global Maltsev, it is straightforward to adopt that algorithm to work in our setting. We can also use the Algorithm in \cite{arash-maltsev}.

\paragraph{Not-Minority Instances} 
\Call{Not-Minority}{} algorithm (Algorithm \ref{alg-not-minority})
first checks whether the instance is a minority instance, and if the answer is yes then it calls \Call{RemoveMinority}{} function. Otherwise, it starts with a not-minority pair $(b,a)$ in $L(x)$, i.e., $w=(x; b^{k}, a)$ with $f(w) = c \ne a$. 
Roughly speaking, the goal is not to use $f$
on vertices $w_1=(x;e_1,e_2,\dots,e_k)$ with $f(w_1)=a$; which essentially means setting $f(w_1)=f(w)$. 
In order to make this assumption, 
it recursively solves a smaller instance of the problem (smaller test), say $G' \subseteq G$, and $L' \subset L$, and if the test is successful then that particular information about $f$ is no longer needed. More precisely, let $w=(x;b^k,a) \in G' \times_{L'} H^k$ so that $f(w)=c \ne a$ and where $(x,a),(x,c)$ are in the same connected component of  $G' \times_{L'} H$. The test $T_{x,c}$ is performed to see whether there exists an $L'$-homomorphism $g$ from $G'$ to $H$ with $g(x)=c$. 
If $T_{x,c}$ for $G',L'$ succeeds then the algorithms no longer uses $f$ for $w'=(x;a_1,a_2,\dots,a_k) \in G' \times_{L'} H^k$ with $f(w')=a$. 
We often use a more restricted test, say $T_{x,c,y,d}$ on $G',L'$ in which the goal is to see whether there exists an $L'$-homomorphism $g$,
from $G'$ to $H$ with $g(x)=c$, $g(y)=d$. 
 
The Algorithm  \ref{alg-not-minority} is recursive, and we use induction on $\sum_{x \in V(G)}|L(x)|$ to show its correctness. In what follows we give an insight of why the weak $k$-NU ($k>2$) property of $H$ is necessary for our algorithm.
For contradiction, suppose $w_1= (x;b^k , a)$ with $f(w_1)=c$ and $w_2=(x;a ,b,b,\dots,b)$ with $f(w_2)=d$. 
If $d=a$ then in \Call{Not-Minority}{} algorithm we try to remove $a$ from $L(x)$ (not to use $a$ in $L(x)$) 
if we start with $w_1$ while we do need to keep $a$ in $L(x)$ because we 
later need $a$ in $L(x)$ for the Maltsev algorithm. It might be the case that $d \ne a$ but during 
the execution of Algorithm \ref{alg-not-minority} for some $w_3=(x;b^k , e)$ with $f(w_3) \ne e$ we assume $f(w_3)$ is $e$. So we need to have $f(w_1)=f(w_2)$, the 
weak NU property, to start in Algorithm \ref{rec-alg-main}.

\begin{algorithm}[ht]
    \caption{ ruling out not-minority pairs, $f$ remains a homomorphism of $G \times_L H^k$ to $H$  and for every $x \in V(G)$, $a',b' \in L(x)$, we have $f(x;a'^k,b')=b'$}
  \label{alg-not-minority}
    \begin{algorithmic}[1]
   
   \State \textbf{Input:} Digraphs $G,H$, lists $L$ and, a weak NU homomorphism $f : G \times_L H^k \rightarrow H$ 
   
     \Function{Not-Minority}{$G,H,L,f$}
     
      \If { $\forall$ $x \in V(G)$, $|L(x)|=1$ }
         $\forall$ $x \in V(G)$ set $g(x)=L(x)$ 
       \State {\textbf {return}} $g$
      \EndIf 
       
   \State If $G \times_L H$ is not connected then consider each  connected component separately 
   \If { all the pairs are minority  } \label{line5}
     \State $g =$ \Call{RemoveMinority}{$G,H,L,f$}
     \State \textbf {return} $g$  \label{line7}
   
   \EndIf 
    
   
   \ForAll {$y,z \in V(G)$, $d_1,d_2 \in L(y)$, $e_1 \in L(z)$ s.t.
     $(d_1,e_1),(d_2,e_1) \in L(y,z)$ }
    \State $(G',L')=$\Call{Sym-Diff}{$G,L,y,d_1,d_2, z,e_1$} 
    \State $g^{y,z}_{d_1,e_1}=$ \Call{Not-Minority}{$G',H,L',f$} 
    \If { $g^{y,z}_{d_1,e_1}$ is empty }  
          
          \State Remove $(d_1,e_1)$ from $L(y,z)$, and remove $(e_1,d_1)$ from $L(z,y)$ 
          \If {$L(y,z)$ is empty} {\textbf {return}  $\emptyset$}
             \EndIf
     \Else \If {$G'=G$}  
      {\textbf {return}  $g^{y,z}_{d_1,e_1}$ }
     \EndIf 
    \EndIf
    \EndFor

    \State $(L,f)=$\Call{Bi-Clique-Instances}{$G,H,L,f$}

    \State \Call {PreProcessing}{$G,H,L$} 
    \If { $\exists$ empty list or $\exists$ empty pair list}   {\textbf {return}} $\emptyset$ 
     \Else {}

    \State $g =$ \Call{RemoveMinority}{$G,H,L,f$}
      
     \State {\textbf {return}} $g$
      
    \EndIf
   \EndFunction 
  \end{algorithmic}

 \end{algorithm}


\noindent {\em For implementation,} 
we update the lists $L$ as well as the pair lists, depending on the output of $T_{x,c}$. If $T_{x,c}$ fails (no $L$-homomorphism from $G$ to $H$ that maps $x$ to $c$) then we remove $c$ from $L(x)$ and if $T_{x,c,y,d}$ fails (no  $L$-homomorphism from $G$ to $H$ that maps $x$ to $c$ and $y$ to $d$)  then we remove $(c,d)$ from $L(x,y)$.
The \Call{Not-Minority}{} takes $G,L,f$ and checks whether all the lists are singleton, and in this case the decision is clear. It also handles each connected component 
of $G \times_L H$ separately.  If all the pairs are minority then it calls \Call{RemoveMinority}{} which is essentially checking for a homomorphism when the instance admits a Maltsev polymorphism.  Otherwise, it proceeds with function \Call{Sym-Diff}{}.

\paragraph{Sym-Diff function} 
Let $y,z \in V(G)$ and $d_1,d_2 \in L(y)$ and $e_1 \in L(z)$ such that $(d_1,e_1),(d_2,e_1) \in L(y,z)$. Let $L_1=L_{z,e_1}$. We consider the instances $G',H,L',f$ of the problem as follows. 
The induced sub-digraph $G'$ consists of vertices $v$ of $G$ such that for every $(d_1,j) \in L_1(y,v)$ we have $(d_2,j) \not\in L_1(y,v)$. Now for each $v \in G'$, $L'(v)=\{i \mid (d_1,i) \in L_1(y,v)\}$. Such an instance is constructed by 
function \Call{Sym-Diff}{$G,L,y,d_1,d_2,z,e_1$}. 

Let $B(G')$ denote a set of vertices $u$ in $G \setminus G'$ that are adjacent (via an out-going or in-coming arc) to some vertex $v$ in $G'$. $B(G')$ is called the boundary of $G'$. We do not add $B'(G')$ into $G'$ which makes it easier to argue about running time. However, we some extra care it is also possible to have an algorithm that include the boundary vertices. 
Note that for every $v \in V(G')$, $|L'(v)| < |L(v)|$. Moreover, $L'(z)=\{e_1\}$, and $L'(y)=\{d_1\}$.   

For $y,z \in V(G)$ and $d_1,d_2 \in L(y)$ and $e_1 \in L(z)$ we solve the instance $G',H,L',f$, by calling the \Call{Not-Minority}{$G',H,L',f$} function. The output of this function call is either a non-empty $L'$-homomorphism $g^{y,z}_{d_1,e_1}$ from $G'$ to $H$ or there is no such homomorphism. If $g^{y,z}_{d_1,e_1}$ does not exist 
then there is no homomorphism from $G$ to 
$H$ that maps $y$ to $d_1$, and $z$ to $e_1$. In this case we remove $(d_1,e_1)$ from $L(y,z)$, and remove $(e_1,d_1)$ from $L(z,y)$. This should be clear because $G'$ is an induced sub-digraph of $G$, and for every vertex $v \in V(G') \setminus \{y,z\} $, $L'(v,z)=L(v,z)$, $L'(v,y)=L(v,y)$. Moreover, it is easy to see (will be shown later) that $L'$ is closed under $f$. The  homomorphism $g^{y,z}_{d_1,e_1}$ is used in the correctness proof of the Algorithm \ref{alg-not-minority}. 

\paragraph{Implementation Remark:} In order to avoid several redundant tests, when a sub-digraph $G'$ constructed by \Call{Sym-Diff}{$G,L,y,d_1,d_2,z,e_1$} is the same as $G$; and the test returns true, then we no longer need to run the test for any other sub-instance $(G'',L'')$ because we just need to return yes to the parent sub-routine calling the current sub-routine.

 \begin{algorithm}[ht]

 \begin{algorithmic}[1]
  \State \textbf{Input:} Digraphs $G$, lists $L$ and, $y,z \in V(G)$, $d_1,d_2 \in L(y)$, $e_1 \in L(z)$ 
 \Function{Sym-Diff} {$G,L,y,d_1,d_2, z,e_1$}
     
    %
     
     \State Create new empty lists $L'$ and  let  
     $L_1=L_{z,e_1}$, and construct the pair lists $L_1 \times L_1$ from $L_1$

      \State Set $L'(z)=e_1$, $L'(y)=d_1$, and set $G'=\emptyset$

      \ForAll { $v \in V(G)$ s.t. $\forall i$ with $(i,d_1) \in L_1(v,y)$ we have $(i,d_2) \not\in L_1(v,y)$ } \label{line12}
      \State add $v$ into set $G'$

     \EndFor  
     \State Let $G'$ be the induced sub-digraph of $G$

    \State Let $B(G')=\{u \in V(G \setminus G') \mid \text{ $u$ is adjacent(in-neighbor or out-neighbor) to  some $v \in V(G')$} \}$

     
     \ForAll {$ u \in V(G')$} 
     
      \State $L'(u)=\{ \text {$i \mid   (d_1,i) \in L_1(y,u) $} \}$ 
     
     \EndFor
     \ForAll { $u,v \in V(G')$  }
       
\State $L'(u,v)=\{ (a,b) \in L_1(u,v) |  (a,b) \ \ is \ \ consistent \ \ in \ \ G' \} $.

     \EndFor

    \State {\textbf {return}} $(G',L')$

 \EndFunction

 \end{algorithmic}

 \end{algorithm}

\begin{figure}[H]
   \begin{center}
 \includegraphics[scale=0.5]{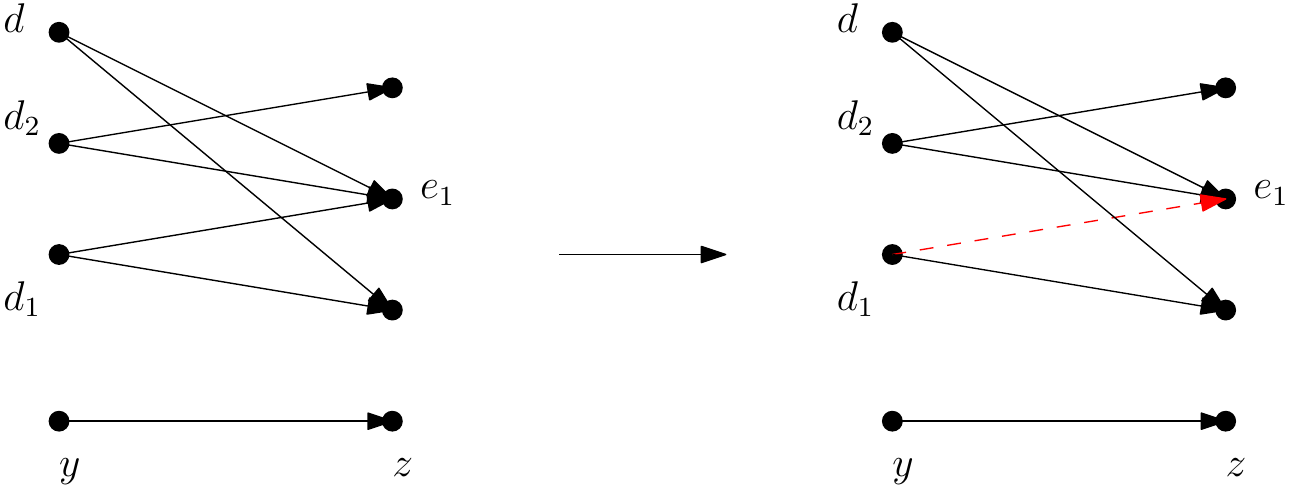}
   \end{center}
   \caption{ Explanation of how to reduce the pair lists. 
   }
 \label{fig:Bi-Clique-Intro}
  \end{figure}

\paragraph{Bi-clique Instances} 
When instance $I=G \times_{L} H$ has more than one connected component we consider each connected component separately. 
Otherwise, there exists a vertex $y$ and two elements $d_1,d_2 \in L(y)$ along with $z \in V(G)$, $e_1 \in L(z)$ such that $(d_1,e_1),(d_2,e_1) \in L(y,z)$. 
Let $d=f(y;d_2^k,d_1) \ne d_1$. 
Suppose $(d_1,e_1),(d_2,e_1) \in L(y,z)$ after running \Call{Not-Minority}{} on the instances from \Call{Sym-Diff}{$G,L,y,d_1,d_2,z,e_1$} and  \Call{Sym-Diff}{$G,L,y,d_2,d_1,z,e_1$}. 
Then we remove $(d_1,e_1)$ from $L(y,z)$ (see Figure \ref{fig:Bi-Clique-Intro}). We continue this until all the pairs are minority. 

After the \Call{Bi-clique-Instances}{} function, we update the lists $L$ by calling PreProcessing, because reducing the pair lists may imply to remove some elements from the lists of some elements of $G$. 
At this point, if $a \in L(x)$ then $f(x;a,a,\dots,a)=a$. This is because when $a$ is in $L(x)$ then it means the \Call{Not-Minority}{} procedure did not consider $a$ to be excluded from further consideration. Notice that in order to feed this instance to \Call{RemoveMinority}{} we only need the idempotent property for those vertices that are in $L(x)$, $x \in V(G)$.

\begin{algorithm}

 \begin{algorithmic}[1]
  \State \textbf{Input:} Digraphs $G$, lists $L$ and, a weak NU homomorphism $f : G \times_L H^k \rightarrow H$ 
 \Function{Bi-Clique-Instances} {$G,H,L,f$ }
 \label{Bi-clique}     
    
 \State update=true
 \While {update }
 
  \State update=false
  \State Let $ d_1,d_2 \in L(y)$, and $e_1 \in L(z)$ such that $(d_1,e_1),(d_2,e_1) \in L(y,z)$ and $f(y;d_2^k,d_1) \ne d_1$

        \If { there is such $y,z,d_1,d_2,e_1$ }

       
        \State Remove $(d_1,e_1)$ from $L(y,z)$ and remove $(e_1,d_1)$ from $L(z,y)$

         \State \Call {PreProcessing}{$G,H,L$} 
     
          \State update=true

      \EndIf


    \EndWhile

    \State {\textbf {return}} $(L)$
    
 \EndFunction

 \end{algorithmic}

 \end{algorithm}


\newpage

\subsection{Examples}\label{example}

\begin{figure}[H]
  \begin{center}
   \includegraphics[scale=0.65]{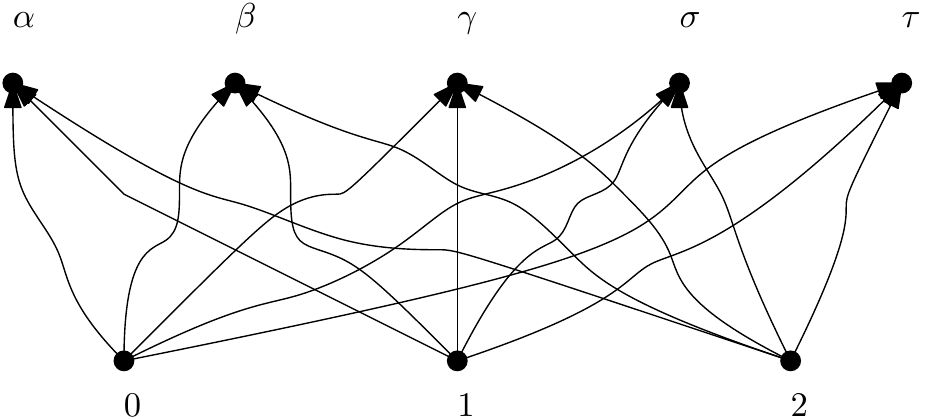}
  \end{center}
  \caption{ \small{There are oriented paths from $0,1,2$ to any of $\alpha,\beta,\gamma,\sigma,\tau$ }}
\label{fig:H-digraph}
 \end{figure}

\begin{figure}
  \begin{center}
   \includegraphics[scale=0.65]{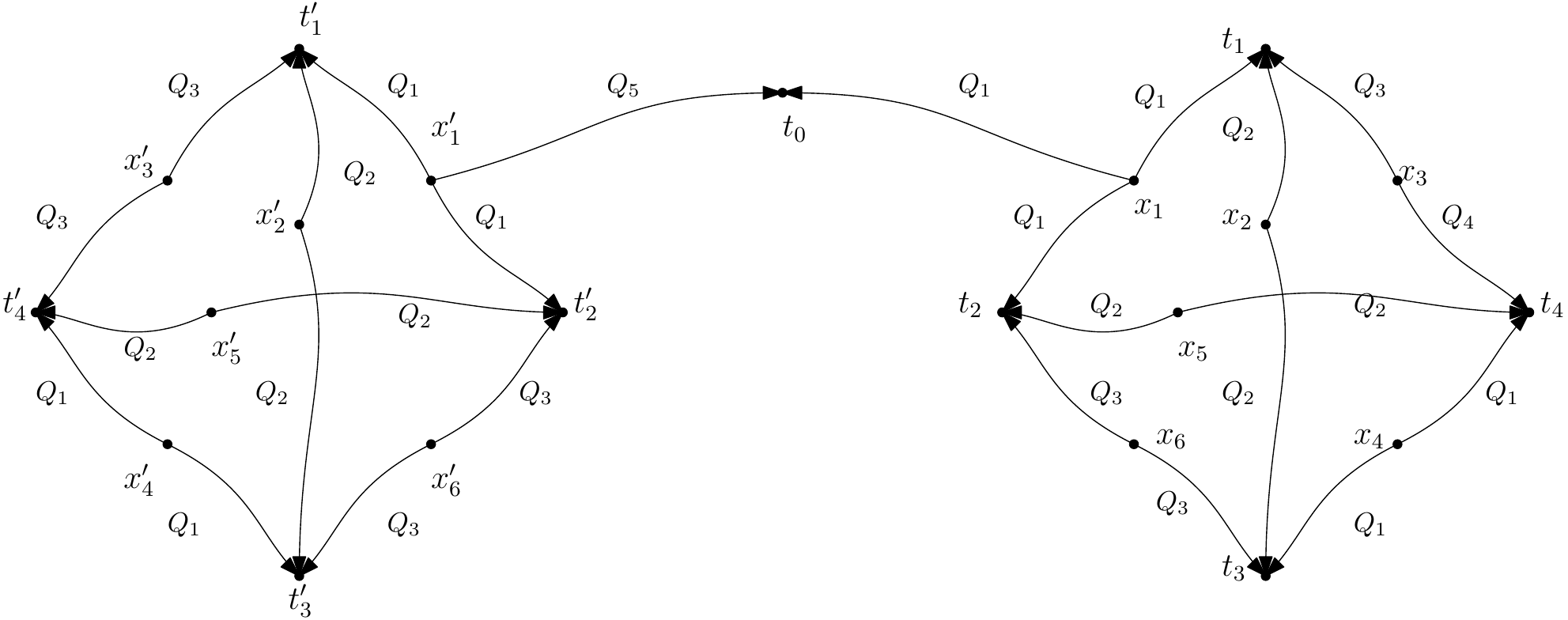}
  \end{center}
  \caption{ The input digraph $G$ }
\label{fig:G-digraph}
 \end{figure}

We refer the reader to the example in \cite{W17}. The digraph $H$ (depicted in Figure \ref{fig:H-digraph}) admits a weak NU of arity $3$. The input digraph $G$ ( depicted in Figure \ref{fig:G-digraph}). This digraph contains the original digraph used in \cite{W17} as an induced sub-digraph. So, our example here combines both examples used in \cite{W17}.  

\begin{figure}[H]
  \begin{center}
   \includegraphics[scale=0.70]{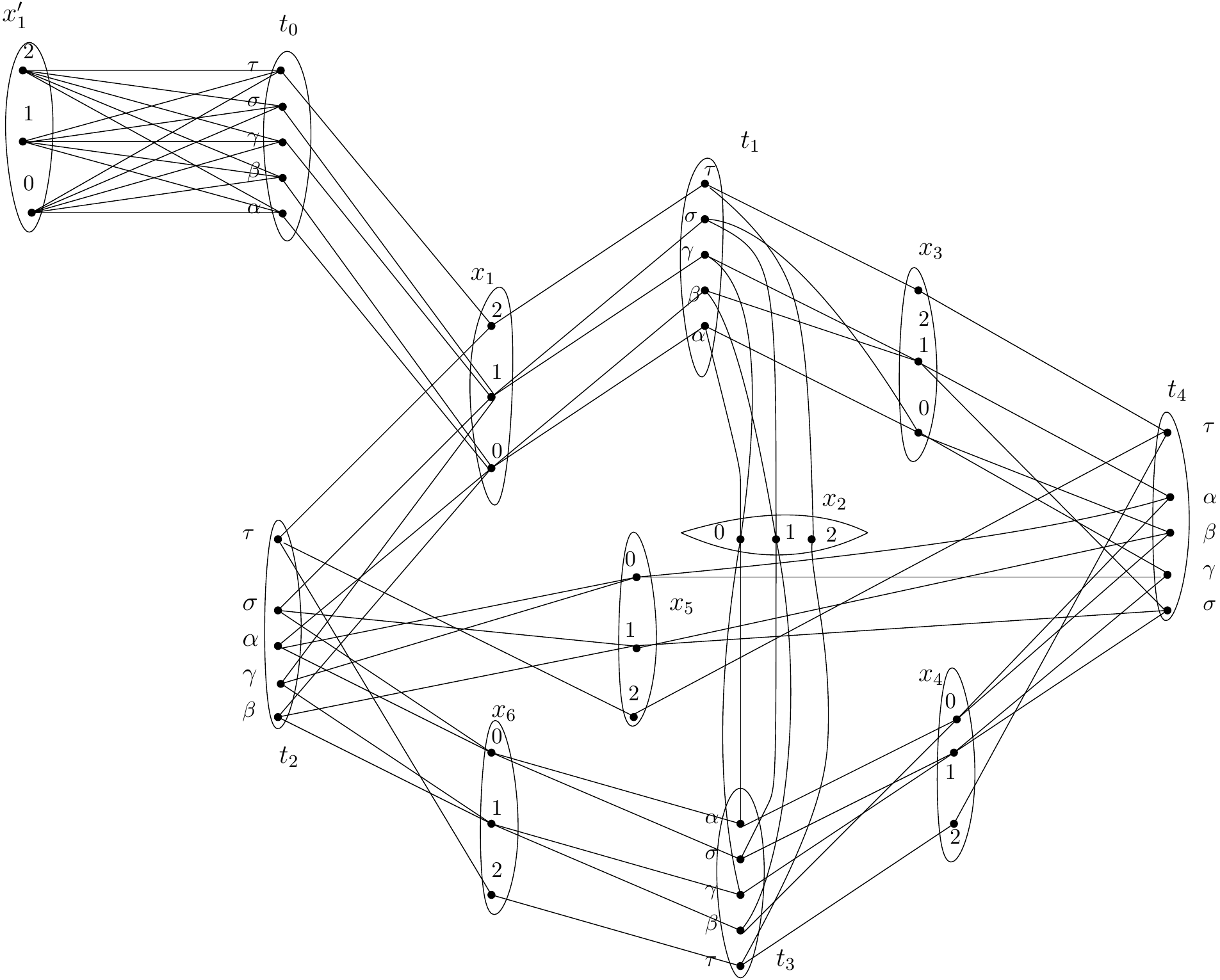}
  \end{center}
  \caption{ \small{Vertices in the ovals show the elements of $H$ that are in the list  of the vertices in $G$, the oriented path  between them are according to path $Q_1,Q_2,Q_3,Q_4$ in $G$. For example, if $x_1$ is mapped to $0$ then $t_1$ can be mapped to $\alpha,\beta$.}}
\label{fig:Ross-example-1}
 \end{figure}

The lists in $G \times H$ are depicted in Figure \ref{fig:Ross-example-1}. 
Clearly the $G \times_L H$ has only one weakly connected component.

\begin{figure}[H]
  \begin{center}
   \includegraphics[scale=0.55]{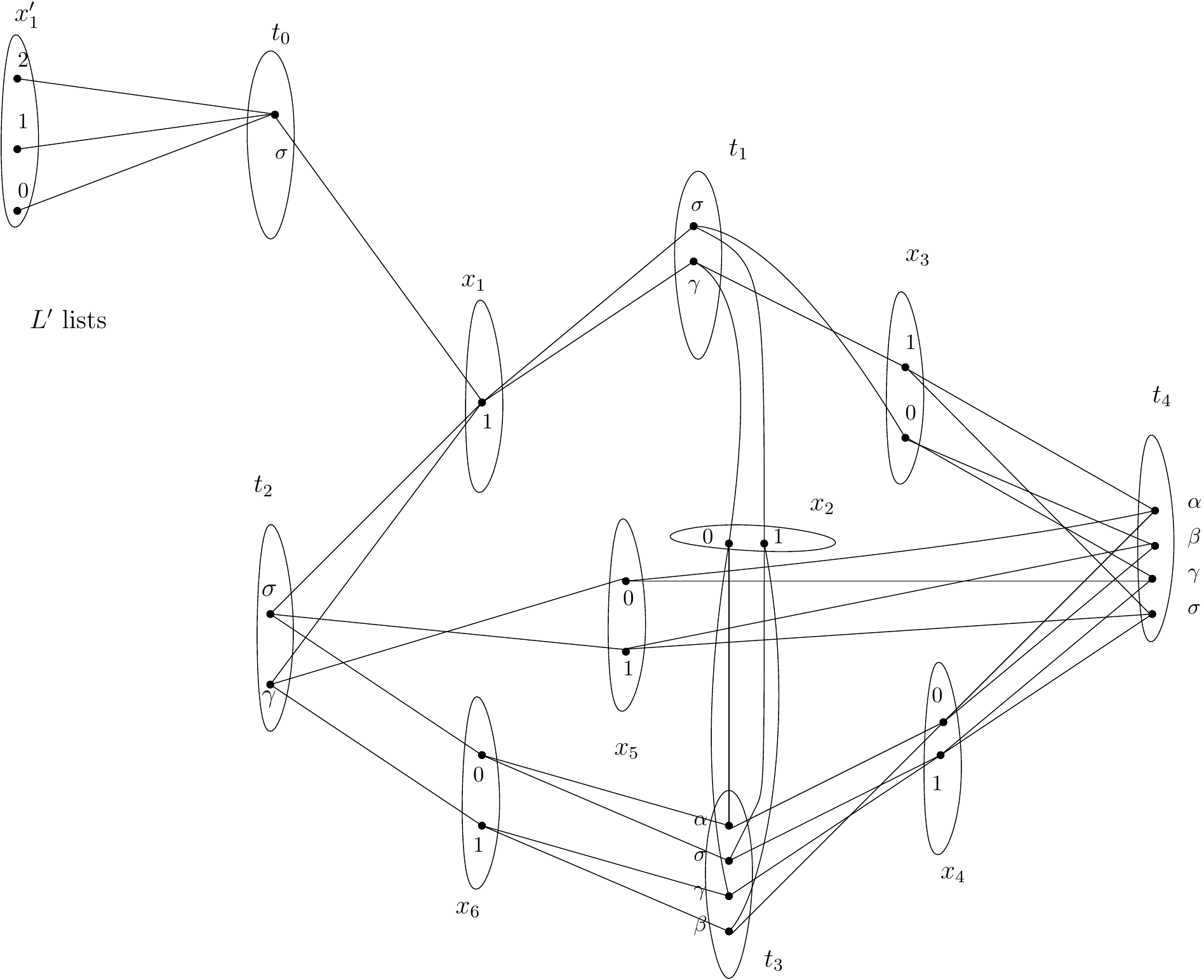}
\end{center}
\caption{ \small{Applying Sym-Diff function on $G,L,t_0,\alpha,\tau,x'_1,1$ and get lists $L'$  }}
\label{fig:Ross-example-2-up}
 \end{figure}

Now consider the \Call{Sym-Diff}{$G,L,t_0,\sigma,\tau,x'_1,1$}. According to the construction of \Call{Sym-Diff}{} we have, $V(G')=\{x'_1,t_0,x_1,x_2,x_3,x_4,x_5,x_6,t_1,t_2,t_3,t_4\}$. This is because $(\tau,1),(\sigma,1) \in L(t_0,x'_1)$, and hence, $G'$ is not extended to  the vertices $\{x'_2,x'_3,x'_4,x'_5,x'_6, t'_1,t'_2,t'_3,t'_4\}$. The list of the vertices in the new instance ($L'$ lists) are as follows. $L'(x'_1)=\{1\}$, $L'(t_0)=\{\sigma\}$, $L'(t_1)=\{\sigma,\gamma\}$, $L'(t_2)=\{\sigma,\gamma\}$, $L'(x_1)=\{1\}$, $L'(x_2)=L'(x_3)=L'(x_4)=L'(x_5)=L'(x_6)=\{0,1\}$, and  $L'(t_3)=L'(t_4)=\{\alpha,\beta,\gamma, \sigma\}$ (See Figure \ref{fig:Ross-example-2-up}).

\begin{figure}[H]
  \begin{center}
   \includegraphics[scale=0.60]{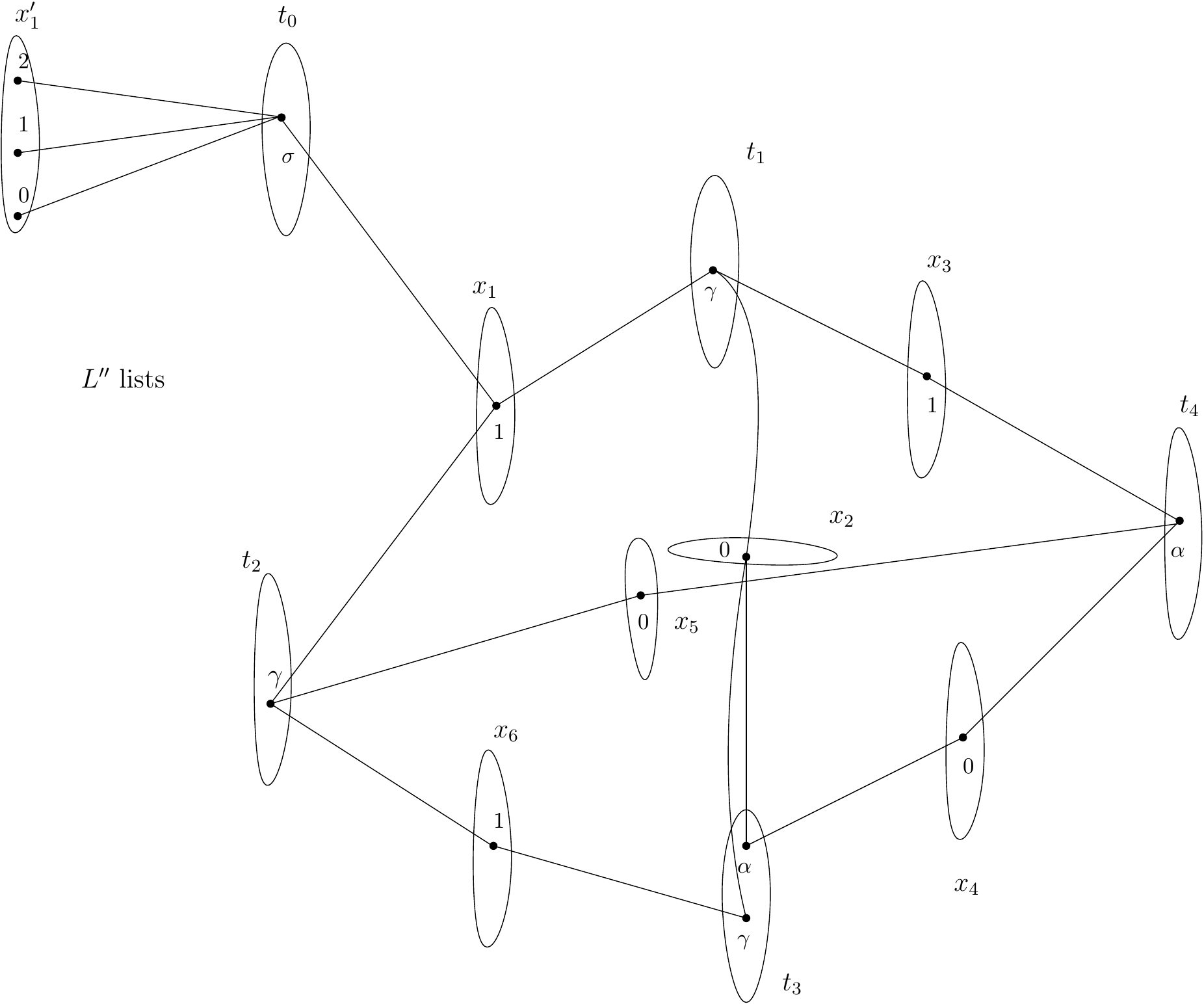}
  \end{center}
  \caption{ \small{Applying Sym-Diff function on $G',L',t_4,\alpha,\beta,x_4,0$}}
\label{fig:Ross-example-2-down}
 \end{figure}

Now suppose we want to solve the instance $G',L'$. At this point one could say this instance is Minority and according to Algorithm \ref{alg-not-minority} we should call a Maltsev algorithm. However, we may further apply Algorithm \ref{alg-not-minority} on instances constructed in \Call{Sym-Diff}{} and use the "no" output to decide whether there exists a homomorphism or not. Now consider  \Call{Sym-Diff}{$G',L',t_4,\alpha,\beta,x_4,0$} (see Figure \ref{fig:Ross-example-2-down}), we would get the digraph $G''=\{x_1,x_2,x_3,x_4,x_5,x_6\}$ and lists $L''$ as follows. 
$L''(t_4)=\{\alpha\}$, $L''(x_4)=\{0\}$. By following $Q_4$ from $t_4$ to $x_3$ we would have $L''(x_3)=\{1\}$ (because $(0,\alpha) \not\in L'(x_3,t_4)$), and 
then following $Q_1$ to $t_1$ we would have $L''(t_1)=\{\gamma\}$ (because $(1,\sigma) \not\in L'(x_3,t_1)$). By similar reasoning and following $Q_2$ from $t_1$ to $x_2$ we have $L''(x_2)=\{0\}$, and consequently from $x_2$ to $t_3$ alongside $Q_2$ we have $L''(t_3)=\{\alpha, \gamma\}$. By following $Q_2$ from $t_4$ to $x_5$ we would have $L''(x_5)=\{0\}$, and consequently $L''(t_2)=\{\gamma\}$. By following $Q_3$ from $t_2$ to $x_6$ we have $L''(x_6)=\{1\}$, and continuing along $Q_3$ to $t_3$ we would have $L''(t_3)=\{\gamma\}$. 

Finally we see the pair lists $L''(x_4,x_6)=\emptyset$ because $(0,1) \not\in L''(x_4,x_6)$. Therefore, there is no $L''$-homomorphism from $G''$ to $H$. This means in the algorithm we remove 
$(\alpha,0)$ from $L'(t_4,x_4)$. Moreover, $L'(\alpha,1) \not\in L'(t_4,x_4)$ and hence $\alpha$ is removed from $L''(t_4)$. Similarly we conclude that there is no homomorphism for the instance  \Call{Sym-Diff}{$G',L',t_4,\beta,\alpha,x_4,0$}, and hence, we should remove $\beta$ from $L'(t_4)$. 

Again suppose we call,  \Call{Sym-Diff}{$G',L',t_4,\gamma,\sigma,x_4,1$}. We would get digraph $D_1$ with vertex set $\{x_1,x_2,x_3,x_4,x_5,x_6\}$ and lists $L_1$ as follows. $L_1(t_4)=\{\alpha\}$, $L_1(x_4)=\{1\}$, $L_1(x_3)=\{0\}$, $L_1(t_1)=\{\sigma\}$, $L_1(x_2)=\{1\}$. $L_1(x_5)=\{0\}$, $L_1(t_2)=\{\gamma\}$, $L_1(x_6)=\{1\}$, and $L_1(t_3)=\{\beta\}$. 
Now we see the pairs lists $L_1(t_3,x_4)=\emptyset$ because $(\beta,1) \not\in L'(t_3,x_4)$. Therefore, we conclude $\gamma$ should be removed from $L'(t_4)$. By similar argument, we conclude that $\sigma$ is removed from $L'(t_4)$.

In conclusion,  there is no $L$-homomorphism that maps $x_1$ to $1$. By symmetry we conclude that there is no $L$-homomorphims that maps $x_1$ to zero. This means $L(x_1)=L(x_2)=L(x_3)=L(x_4)=L(x_5)=L(x_6)=\{2\}$. $L(t_1)=L(t_2)=L(t_3)=L(t_4)=\{\tau\}$. So any homomorphism $\phi$ from $G$ to $H$ maps $x_i$, $1 \le i \le 6$ to $2$ and it maps $t_i$, $1 \le i \le 4$ to $\tau$. 

It is easy to verify that $\phi$  may map all the vertices $x'_1,x'_2,x'_3,x'_4,x'_5,x'_6$ to $2$, and there also exists a  homomorphism $\phi'$ where the  image of  $x'_1,x'_2,x'_3,x'_4,x'_5,x'_6$ is in $\{0,1\}$. 


 
\paragraph{Generalization} 
Let $R$ be a relation of arity $k$ on set $A$, and suppose $R$ admits a weak NU polymorphism $\phi$ of arity $3$ (for simplicity). 
Let $\alpha_1,\alpha_2,\dots,\alpha_m$ be the tuples in $R$. Let $a_1,a_2,\dots,a_n$ be the elements of $A$. Let $\alpha_j=(c_1,c_2,\dots,c_k)$ be the $j$-tuple, $1 \le j \le m$ of $R$. 
Let $P_{i,j}$ be an oriented path that is constructed by concatenating  $k+2$ smaller pieces (oriented path) where each piece is either a forward arc or a forward-backward-forward arc. The first piece of $P_{i,j}$ is a forward arc and the $k+2$-piece is also a forward arc. The $r$-th piece, $2 \le r \le k$, is a forward arc if $a_i=c_r$, otherwise, the $r$-th piece is a forward-backward-forward arc; and in this case we say the $r$-th piece has two internal vertices. Note that $(r+1)$-the piece is attached to the end of the $r$-th piece.  For example, if $a_1=0$ and $\alpha_1=(0,0,0,1,0)$ then $P_{1,1}$ looks like :

\begin{figure}[H]
  \begin{center}

   \includegraphics[scale=0.70]{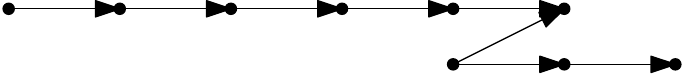}
  \end{center}
  \caption{ The oriented path corresponding to $(0,0,0,1,0)$}
\label{fig:zig-zag}
 \end{figure}
%
Now $H$ is constructed as follows. $V(H)$ consists of $b_1,b_2,\dots,b_n$ corresponding to $a_1,a_2,\dots,a_n$,  together with vertices $\beta_1,\beta_2,\dots,\beta_m$ corresponding to $\alpha_1,\alpha_2,\dots,\alpha_m$. For every
$1 \le i \le n$, $1 \le j \le m$, we put a copy of $P_{i,j}$ between the vertices $b_i$ and $\beta_j$; identifying the beginning of $P_{i,j}$ with $b_i$ and end of $P_{i,j}$ with $\beta_j$. 

Now it is easy to show that the resulting digraph $H$ is a balanced digraph and admits a weak NU polymorphism. For every triple $(b_i,b_j,b_{\ell})$ from $b_1,b_2,\dots,b_n$ set $\psi(b_i,b_j,b_{\ell})=b_s$ where $a_s=\phi(a_i,a_j,a_{\ell})$. For every 
$\beta_{i},\beta_{j},\beta_{\ell}$ from $\{\beta_1,\beta_2,\dots,\beta_m\}$, set $\psi(\beta_{i},\beta_{j},\beta_{\ell})=\beta_{s}$ where  $\alpha_s=\phi(\alpha_i,\alpha_j,\alpha_{\ell})$ where $\phi$ is applied coordinate wise on $(\alpha_i,\alpha_j,\alpha_{\ell})$. We give level to the vertices of $H$. All the vertices, $b_1,b_2,\dots,b_n$ gets level zero. If $uv$ is an arc of $H$ then $level(v)=1+level(u)$. All the $\beta_1,\beta_2,\dots,\beta_m$ vertices gets level $k+2$. For every $a,b,c \in V(H)$, set $\psi(a,b,c) =a$  when  
$a,b,c$ are not on the same level of $H$. Otherwise, for $a \in P_{i,i'}$ and $b \in P_{j,j'}$, and  
$c \in P_{\ell,\ell'}$ where all on level  $h$ of $H$, set $\psi(a,b,c)= d$ where $d$ has the following properties:
\begin{itemize}
\item $d$ is a vertex on the same level as $a,b,c$,
\item $d$ lies on $P_{s,s'}$ where $\phi(a_i,a_j,a_{\ell})=a_s$ and $\phi(\alpha_{i'},\alpha_{j'},\alpha_{\ell'})=\alpha_{s'}$,
\item if any of the $a \in P_{i,i'}$, $b \in P_{j,j'}$, and  
$c \in P_{\ell,\ell'}$ is an interval vertex (reffering to the r-th piece of $P$) then $d$ is also an interval vertex of $P_{s,s'}$ when exists. Otherwise $d$ should not be an internal. 

\item if $i=j=\ell$ and $i'=j'=\ell'$, then $\psi(a,b,c)=a$ if $a=b$ or $a=c$, otherwise,  $\psi(a,b,c)=b$ (i.e., the majority function). 
\end{itemize}

Suppose $aa',bb',cc'$ are arcs of $H$. By the following observation, it is easy to see that $\psi(a,b,c)\psi(a',b',c')$ is an arc of $H$. 

\noindent {\em Observation.} Suppose  $a,b,c$ are at the beginning (end) of the $r$-th piece of $P_{i,i'}$ and $b \in P_{j,j'}$, and  $c \in P_{\ell,\ell'}$ (respectively) and none of these three pieces has an interval vertex. Then,  the $r$-th piece of $P_{s,s'}$ does not an interval vertex. 

By definition $a_i$ appears in the $r$-th coordinate of $\alpha_{i'}$, and $a_j$ appear in the $r$-th coordinate of $\alpha_{j'}$, and $a_{\ell}$ appears in the $r$-th coordinate of $\alpha_{\ell'}$. Since $\phi$ is applied coordinate wise on $(\alpha_{i'},\alpha_{j'},\alpha_{\ell'})$, the $r$-th coordinate of $\alpha_{s'}$ is $\phi(a_i,a_j,a_{\ell})=a_s$, and hence, the $r$-th coordinate of $\alpha_{s'}$ is $a_s$. Therefore, the $r$-th piece of $P_{s,s'}$ doesn't have an interval vertex.

\section{Proof of Theorem \ref{main-theorem} }\label{sec:proofs}


By Lemma \ref{rec-apf-preserve},  we preserve the existence of a homomorphism from 
$G$ to $H$ after Algorithm. We observe that the running time of PreProcessing function is $\mathcal{O}(|G|^3|H|^3)$.  According to the proof of Lemma \ref{rec-apf-preserve} (2) the running time of Algorithm \ref{alg-not-minority} is 
$\mathcal{O}(|G|^4 |H|^{k+4})$. 
Therefore, the running time of the Algorithm \ref{rec-alg-main} is $\mathcal{O}(|G|^{4} |H|^{k+4})$.

\subsection{PreProcessing and List Update}

We first show that the standard properties of consistency checking remain true in our setting -- namely, that if the Preprocessing algorithm succeeds then $f$ remains a homomorphism consistent with the lists $L$ if it was before the Preprocessing.

\begin{lemma} \label{rec-k-tuple-consistency}
If $f$ is a homomorphism of $G \times H^k \rightarrow H$ consistent with $L$ then $f$ is a homomorphism consistent with $L$ after running the Preprocessing.
\end{lemma}
\pf We need to show that if $a_1,a_2,\dots,a_k$ are in $L(y)$ after the Preprocessing then \\
$f(y;a_1,a_2,\dots,a_k) \in L(y)$ after the Preprocessing. 
By definition vertex $a$ is in $L(y)$ after the
Preprocessing because for every oriented path $Y$ (of some length $m$) in $G$ from $y$ to a fixed vertex $z \in V(G)$
there is a vertex $a' \in L(z)$ and there exists a walk $B$ in $H$ from $a$ to $a'$ and congruent with $Y$ that lies in $L(Y)$; list of the vertices of $Y$. Let $a'_1,a'_2,a'_3,\dots,a'_k \in L(z)$. Let $A_i$, $1 \le i \le k$ be a walk from $a_i$ to $a'_i$ in $L(Y)$ and congruent to $Y$. Let $A_i=a_i,a_1^i,a^2_i,\dots,a^m_i,a'_i$
and let $Y=y,y_1,y_2,\dots,y_m,z$. Since $f$ is a homomorphism consistent with $L$ before the Preprocessing, $f(y;a_1,a_2,\dots,a_k),  f(y_1;a_1^1,a_2^1,\dots,a_k^1), \dots,$
$f(y_i;a_1^i,a_2^i,\dots,a_k^i),\dots, f(y_m;a_1^m,a_2^m,\dots,a_k^m),\\ f(z;a'_1,a'_2,\dots,a'_k)$  is a
walk congruent with $Y$. This would imply that there is a walk from
$f(y;a_1,a_2,\dots,a_k)$ to $f(z;a'_1,a'_2,\dots,a'_k)$ congruent with $Y$ in $L(Y)$, 
and hence, $f(y;a_1,a_2,\dots,a_k) \in L(y)$.  \qed

By a similar argument as in the proof of Lemma \ref{rec-k-tuple-consistency} we have the following lemma.

\begin{lemma}\label{rec-k-tuple-consistency2}
If $f$ is a homomorphism of $G \times H^k \rightarrow H$, consistent with $L$ and 
$(a_i,b_i) \in L(x,y)$, $1 \le i \le k$, after Preprocessing then  $(f(x;a_1,a_2,\dots,a_k),f(y;b_1,b_2,\dots,b_k)) \in L(x,y)$ after the Preprocessing.
\end{lemma}

\subsection{Correctness Proof for Not-Minority Algorithm}

The main argument is proving that after \Call{Not-Minority}{} algorithm ( Algorithm  \ref{alg-not-minority}), there still exists a homomorphism from $G$ to $H$ if there was one before \Call{Not-Minority}{} .

\begin{lemma} \label{rec-apf-preserve}
If $(d_1,e_1) \in L(y,z)$ after calling \Call{Not-Minority}{$G_1,H,L_1,f$} where $(G_1,L_1)=$ \Call{Sym-Diff}{$G,L,y,d_1,d_2,z,e_1$}, then set $test_1=true$, otherwise, set $test_1=false$.  If $(d_2,e_1) \in L(y,z)$ after calling \Call{Not-Minority}{$G_2,H,L_2,f$} where $(G_2,L_2)=$ \Call{Sym-Diff}{$G,L,y,d_2,d_1,z,e_1$}, then set $test_2=true$, otherwise, set $test_2=false$. Then the following hold. 
\begin{enumerate}
   \item[$\alpha.$]  If $test_1$ is false then there is no homomorphism from $G$ to $H$ that maps  $y$ to $d_1$ and $z$ to $e_1$. 
    \item[$\beta.$] If $test_2$ is false then there is no homomorphism from $G$ to $H$ that maps $y$ to $d_2$ and $z$ to $e_1$. 
    
    \item[$\gamma.$ ] If both $test_1,test_2$ are true then there exists an $L$-homomorphism from $G_1=G_2$ to $H$, that maps $y$ to $d$ and $z$ to $e_1$ where $f(y;d_2^k,d_1)=d \ne d_1$.  Moreover, \Call{Not-Minority}{} returns an $L'$- homomorphism from $G'$ to $H$ where $(G',L')=$\Call{Sym-Diff}{$G,L,y,d,d_1,z,e_1$}.
    
    \item[$\lambda.$ ] Suppose both $test_1,test_2$ are  true, and $f(y;d_2^k,d_1)=d \ne d_1$. 
    Suppose there exists a homomorphism $g$ from $G$ to $H$ with $g(y)=d_1$ and 
    $g(z)=e_1$. Then there exists a homomorphism $h$ from $G$ to $H$ with  $h(y)=d$ and $h(z)=e_1$. 
    
  \end{enumerate}

\end{lemma}
\pf  We use induction on the $\sum_{x \in V(G)}|L(x)|$. The base case of the induction is when all the lists are singleton, or
when all the pairs are minority. If the lists are singleton then at the beginning of \Call{Not-Minority}{} algorithm we check whether the singleton lists form a homomorphism from $G$ to $H$. If all the pairs are minority ($f(x,a_1^k,a_2)=a_2$ for every $x \in V(G)$, $a_1,a_2 \in L(x)$)
then the function \Call{RemoveMinority}{} inside \Call{Not-Minority}{} algorithm, correctly returns a homomorphism from $G$ to $H$ if there exists one (see lines \ref{line5}--\ref{line7} of Algorithm \Call{RemoveMinority}{} \ref{alg-not-minority}). \\
Therefore, we continue by assuming the existence of some not-minority pairs. Consider the instance $G_1,L_1$ constructed by \Call{Sym-Diff}{$G,L,y,d_1,d_2,z,e_1$} in which  $L_1(y)=\{d_1\}$ and $L_1(z)=\{e_1\}$. \\

\noindent {\textbf {Proof of ($\alpha$)}} 
We first notice that $f$ is closed under $L_1$. Let $v \in G_1$ and suppose $c_1,c_2,\dots,c_k \in L_1(v)$. Thus, we have $(c_1,d_1),(c_2,d_1),\dots,(c_k,d_1) \in L(v,y)$, and $(c_1,e_1),(c_2,e_1),\dots,(c_k,e_1) \in L(v,z)$. Let $P_1=v,v_1,v_2,\dots,v_t,y$ (or $z$) be an arbitrary oriented path from $v$ to $y$ ($z$) in $G_1$. Now $c_0=f(v;c_1,c_2,\dots,c_k),f(v_1;c^1_1,c^1_2,\dots,c^1_k),\dots,f(v_t;c^t_1,c^t_2,\dots,c^t_k),f(y;d_1,d_1,\dots,d_1)=d_1$ where $c^i_1,c^i_2,\dots,c^i_k \in L_1(v_i)$, $1 \le i \le t$, implies a path from $c_0$ to $d_1$, and hence, there exists an oriented path from $c_0$ to $d_1$ in $L(P_1)$ and congruent to $P_1$. This would mean $c_0 \in L_1(v)$, according to \Call{Sym-Diff}{} construction. Observe that $G_1$ is an induced sub-digraph of $G$, and  $\sum_{x \in V(G_1)}|L_1(x)| < \sum_{x \in V(G)}|L(x)|$. Thus, by induction hypothesis (assuming  \Call{Not-Minority}{} returns the right answer on smaller instance) for instance $G_1,L_1,H,f$, there is no homomorphism from $G_1$ to $H$ that maps, $y$ to $d_1$ and $z$ to $e_1$.

For contradiction, suppose there exists an $L$-homomorphism $g$ from $G$ to $H$ (with $g(y)=d_1$, $g(z)=e_1$). Then, for every vertex $v \in V(G_1)$ we have $g(v) \in L(v)$, and $(g(v),d_1) \in L(v,y)$, and $(g(v),e_1) \in L(v,z)$. On the other hand, by the construction in function \Call{Sym-Diff}{}, $L_1(v)$ contains every element $ i \in L(v)$ when
$(i,d_1) \in L(v,y)$ and $(i,e_1) \in L(v,z)$, and consequently $g(v) \in L_1(v)$. However, $g_1 : G_1 \rightarrow H$, with $g_1(u)=g(u)$ for every $u \in V(G_1)$ is a homomorphism, a contradiction to nonexistence of such a homomrphism.

Notice that $(d_1,i) \in L(y,v)$ and $(e_1,i) \in L(z,v)$, in the first call to \Call{Sym-Diff}{}. But, if at some earlier call to \Call{Sym-Diff}{}, we removed $(d_1,i)$ from $L(y,v)$ then by induction hypothesis this decision was a right decision, and hence, $i \ne g(v)$, and consequently is not used for $g$. \\

\noindent {\textbf {Proof of ($\beta$)}} is analogous to proof of $(\alpha)$. \\

\noindent {\textbf {Proof of ($\gamma$)}} Suppose $test_1,test_2$ are true. 
Let $g_1$ be the homomorphism returned by \Call{Not-Minority}{} function for the instance $G_1,H,L_1,f$, and $g_2$ be the homomorphism returned by \Call{Not-Minority}{} function for instance $G_2,H,L_2,f$. (i.e. from \Call{Sym-Diff}{$G,L,y,d_2,d_1,z,e_1$}). By definition $G_1=G_2$. Let $G'$ be the digraph constructed in \Call{Sym-Diff}{$G,L,y,d,d_1,z,e_1$}. Notice that $G'$ is an induced sub-digraph of $G_1$. This is because when $z_1$ is in $B(G_1)$ then there exists some $r \in L(v)$ such that $(d_1,r),(d_2,r) \in L(y,v)$, and since $f$ is closed under $L$, we have $(f(v,d_1^k,d_2),r) \in L(y,v)$. Therefore, $v$ is either inside $B(G')$ or $ v \in V(G) \setminus V(G')$; meaning that $G'$ does not expand beyond $B(G_1)$, and hence, $G'$ is an induced sub-digraph of $G_1$. 

Now for every vertex $y \in V(G')$, set $g_3(y)=f(y;g_1^k(y),g_2(y))$. Since $g_1,g_2$ are also  $L$-homomorphism from $G_1=G_2$ to $H$ and $f$ is a polymorphism, it is easy to see that $g_3$ is an $L$-homomorphism from $G_1$ to $H$, and hence, also an $L$- homomorphism from $G'$ to $H$.  \\

\noindent {\textbf {Proof of ($\lambda$)}} Suppose there exists an $L$-homomorphism 
$g : G \rightarrow H$ with $g(y)=d_1$, 
$g(z)=e_1$. Then we show that there exists an $L$-homomorphism $h : G \rightarrow H$ with $h(y)=d$, $h(z)=e_1$. \\

\noindent {\textbf {Remark:}} The structure of the proof is as follows. In order to prove the statement of the Lemma \ref{rec-apf-preserve}($\lambda$) we use Claim \ref{cl1}. The proof of Claim \ref{cl1} is based on the induction on the size of the lists.

Let $g_1$ be an $L$-homomorphism from $G_1$ to $H$ with $g_1(y)=d_1$ and $g_1(z)=e_1$, and $g_2$ be an $L$-homomorphism from $G_2$ to $H$ with $g_2(y)=d_2$, and $g_2(z)=e_1$. 
 According to $\gamma$, there exists an $L$-homomorphism $g_3=g^{y,z}_{d,e_1}$ form $G'=G_1=G_2$ to $H$, that maps $y$ to $d$ and $z$ to $e_1$. As argued in the proof of $\gamma$, $g_3$ is constructed based on $g_1,g_2$ and polymorphism $f$. We also assume that $g_1$ agrees with $g$ in $G_1$. 
 
\begin{figure}[ht]
  \begin{center}
   \includegraphics[scale=0.53]{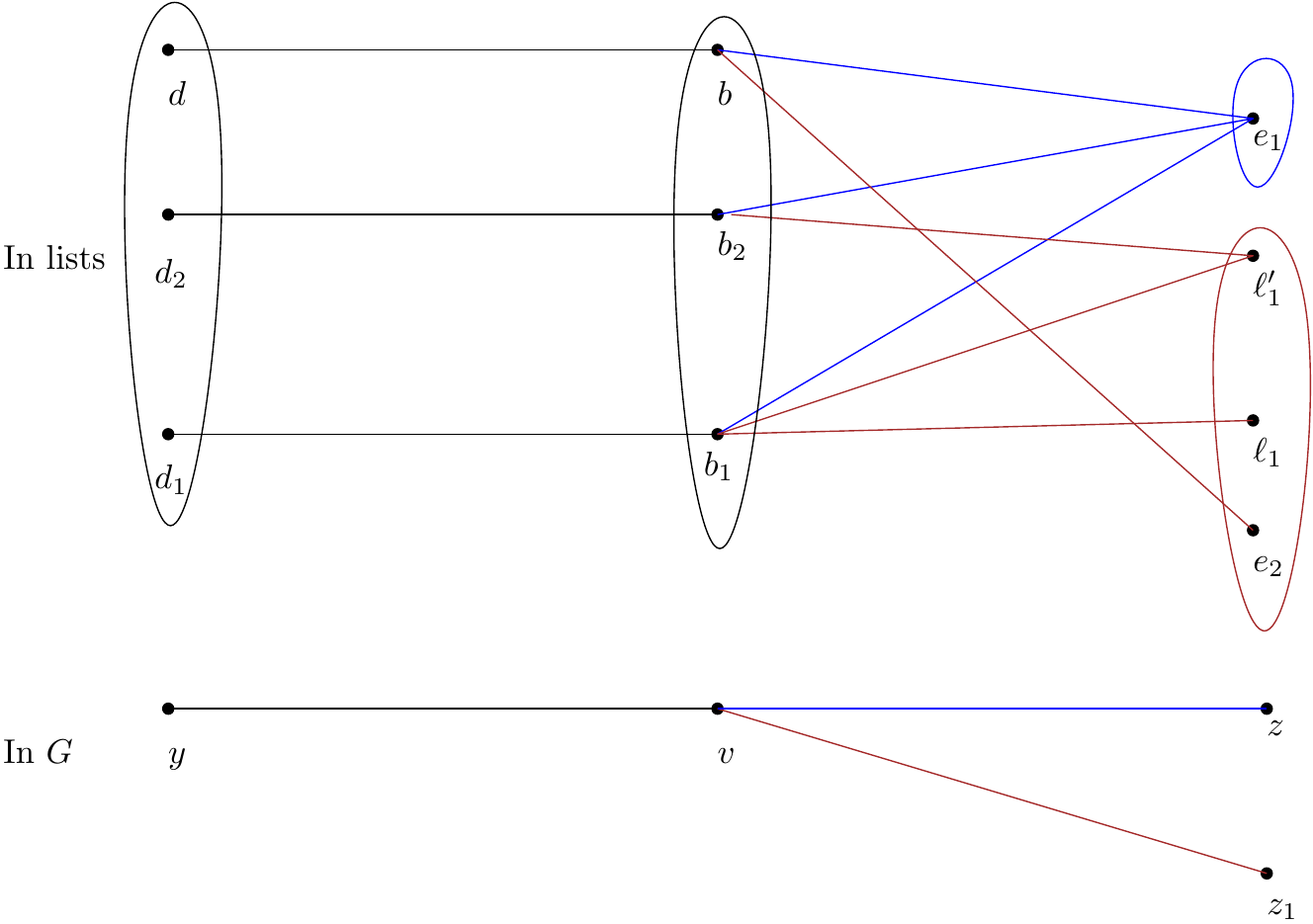}
  \end{center}
  \caption{ \small{$z_1 \in B(G')$, and $\ell_1,\ell'_1 \in L_{z,e_1}(z_1)$, $(b_2,\ell'_1) \in L_{z,e_1,y,d_1}(v,z_1)$, . Thus, there is a walk from $b=f(v;b_2^k,b_1)$ to $e_2=f(z_1;\ell_1'^k,\ell_1)$ in $L(P_1[v,z_1])$ }}
\label{fig:Proof-fig2}
 \end{figure}

 If $G'=G$, then we return the homomorphism $g_3$ as the desired homomorphism. 
Otherwise, consider a vertex $z_1$ which is on the boundary of $G'$, $B(G')$. Recall that $B(G')$ is the set of vertices $u \in V(G)$ with some $i \in L_{z,e_1}(u)$, such that $(i,d_1),(i,d_2) \in L_{z,e_1}(u,y)$. 
Let $P$ be an oriented path in $G' \cup B(G')$ from $y$ to $z$, and let $P_1$ be an oriented path in $G' \cup B(G')$ from $y$ to $z_1$. We may assume $P$ and $P_1$ meet at some vertex $v$ ($v$ could be $y$, see Figure \ref{fig:Proof-fig2}).  
We may assume $z_1$ is chosen such that $\ell_1=g(z_1)$, and $g_3(v_1) \ell_1 \not\in A(H)$ ($\ell_1 g_3(v_1) \not\in A(H)$) where $v_1$ is last vertex  
before $z_1$ on $P_1$, and $v_1z_1  \in A(G)$ ($z_1v_1 \in A(G)$) ($g_3(v_1)$ must have a neighbor inside $L(z_1)$ because of the Preprocessing).  
Notice that if $g_3(v_1)\ell_1 \in A(G)$ for every such path $P_1$ then we can extend $g_3$ to $z_1$ as well. Moreover, 
if there is no such $z_1$ then we can extend $g_3$ to $B(G')$, and define $h(y)=g_3(y)$ for every $y \in V(G')$, and $h(y)=g(y)$ for every $y \in V(G) \setminus V(G')$. It is easy to see that $h$ is an L-homomorphism from $G$ to $H$ with $h(y)=d$. 
Thus, we proceed by assuming the existence of such $z_1$. Let $L_1=L_{y,d_1,z,e_1}$, and $L_2=L_{y,d_2,z,e_1}$. Now we look at $G'$, and the aim is the following. 
\begin{itemize}
    \item First, modify $g$ on the boundary vertices, $B(G')$, so that the image of every $z_i \in B(G')$, $g(z_i) \in L_{y,d,z,e_1}(z_i)$, i.e. $(d,g(z_i)) \in L(y,z_i)$.
    \item Second, having a homomorphism $g_3$ (i.e. $g_3(y)=d$, $g_3(z)=e_1$) from $G' \cup B(G')$ to $H$ that agrees with $g$ on $B(G')$. 
\end{itemize}
Maybe the second goal is not possible on $B(G')$, and hence, we look beyond $B(G')$ and look for differences between $g_3$ and $g$ inside an induced sub-digraph of $G$ containing $G'$. We construct the next part of homomorphism $h$ using $g_3$ and homomrphism $g'_3$ ( obtained from $g$, $g^{z_i,z}_{\ell_i,e_1}$ for some $\ell_i \in L(z_i)$), and homomorphism $f$.
Let $b_1=g(v)$ and let $b_2 =g_2(v)$, and $b=f(v;b_2^k,b_1)$. Let $\ell_1=g(z_1)$, and $\ell'_1 \in L(z_1)$ such that $g_2(v_1) \ell'_1 \in A(H)$ (recall that $v_1z_1$ is the last arc on $P_1$; the path from $y$ to $z$).\\

\noindent {\textbf{First scenario.}}
There exists $\ell'_1 \in L_1(z_1) \cap L_2(z_1)$ such that   
$(b_1,\ell'_1),(b_2,\ell'_1) \in L(v,z_1)$ (see Figure \ref{fig:Proof-fig2}). Note that since $g$ is a homomorphism, we have $(d_1,b_1) \in L(y,v)$ and $(b_1,e_1) \in L(v,z)$. Set $e_2= f(z_1;(\ell'_1)^k,\ell_1). $\\

\noindent {\textbf{Case 1.}} Suppose $e_2 \ne \ell_1$. Let $P_1[v,z_1]=v,v_1,v_2,\dots,v_t,z_1$, and let walk $b_2,c_1,c_2,\dots,c_t,\ell'_1$, and walk $b_1,d_1,\dots,d_t,\ell_1$ be inside $L(P_1[v,z_1])$ and congruent with it. 
Observe that the walk $f(v,b_2^k,b_1),f(v_1,c_1^k,d_1),\dots,f(v_t,c_t^k,d_t),
f(z_1,(\ell'_1)^k,\ell_1)$ inside $L(P_1[v,z_1])$ is from $b=f(v,b_2^k,b_1)$ to $e_2$. Therefore, $(b,e_2) \in L(v,z_1)$ (Figure \ref{fig:Proof-fig2}), and consequently $(d,e_2) \in L(y,z_1)$. \\

\noindent {\textbf{Case 2.}} Suppose $e_2=\ell_1$. Note that in this case again by following the oriented path $P_1[v,z_1]$ and applying the polymorphism $f$ in $L(P_1)$, we conclude that there exists a walk from $b$ to $e_2$ in $L(P_1[v,z_1])$, congruent to $P_1[v,z_1]$, and hence, $(b,\ell_1) \in L(v,z_1)$ and consequently $(d,e_2) \in L(y,z_1)$ \\

Since $G,L_1$ is smaller than the original instance, by Claim \ref{cl1}, the small tests pass for $G,L_1$, and hence, by induction hypothesis, we may 
assume that there exists another $L_1$-homomorphism from $G$ to $H$ that maps $y$ to $d_1$ and $z$ to $e_1$, and $z_1$ to $e_2$. Thus, for the sake of less notations we may assume $g$ is such a homomorphism. This means we may reduce the lists $L_1$ by identifying $\ell_1$ and $e_2$ when $e_2 \ne \ell_1$ in $L_1(z_1)$, and hence, we restrict the lists $L_1$ to $L_1(z_1,e_2)$. Observe that according to Cases 1,2, $(d,e_2) \in L(y,z_1)$.

\begin{figure}[H]
  \begin{center}
   \includegraphics[scale=0.53]{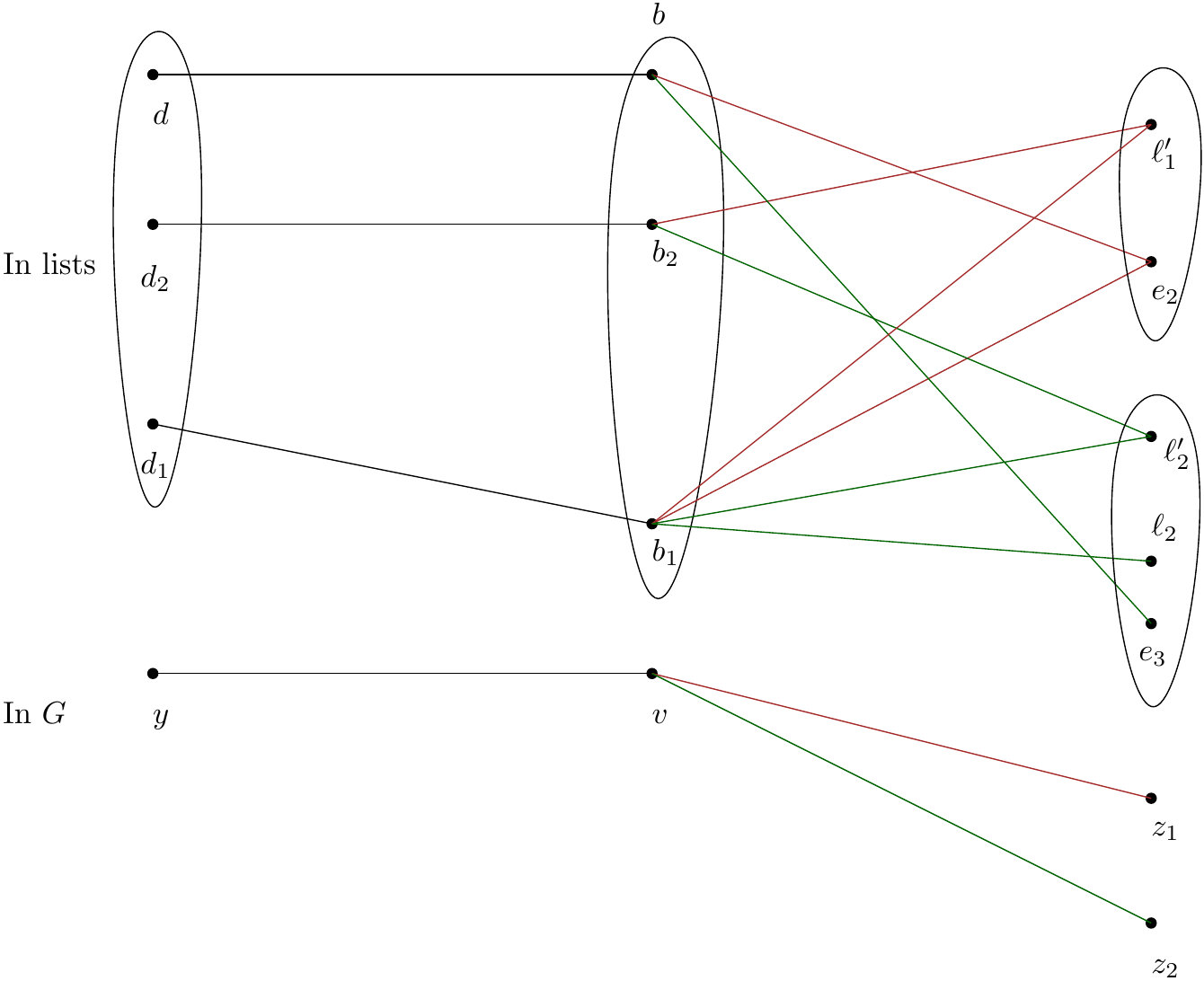}
  \end{center}
  \caption{ \small{$z_2 \in B(G')$, and $\ell_2,\ell'_2 \in L_{z,e_1}(z_2)$ where $\ell_2=g(z_2)$. There is a walk from $b$ to $e_3=f(z_2;(\ell'_2)^k,\ell_2)$}}
\label{fig:Proof-fig3}
 \end{figure}

We continue this procedure as follows: 
Let $z_2$ be the next vertex in $B(G')$ with $g(z_2)=\ell_2$. 
Again if the First scenario occurs, meaning that there exists $\ell'_2 \in L_{2}(z_2)$ such that $(b_2,\ell'_2) \in L(v,z_2)$ and $(b_1,\ell'_2) \in L_1(v,z_2)$ then we continue as follows. Let $e_3=f(z_2; (\ell'_2)^k, \ell_2)$ (see Figure \ref{fig:Proof-fig3}).

If $e_3=\ell_2$ then we have $(b,\ell_2) \in L_1(v,z_2)$ (this is because of the definition of the polymorphism $f$), and hence, as in Case 2 we don't modify $g$. Otherwise, we proceed as in Case 1. In other words, we may assume that there exists an $L_1$-homomorphism from $G$ to $H$ that maps $y_1$ to $d_1$, $z$ to $e_1$ and $z_1$ to $e_2$, and $z_2$ to $e_3$. We may assume $g$ is such a homomorphism. Again this means we further restrict $g$ on the boundary vertices of $G'$; $B(G')$, so they are simultaneously reachable from $b$. If all the vertices in $B(G')$ fit into the First scenario then we return homomorphism $h$ from $G$ to $H$ where inside $G'$, $h$ agrees with $g_3$ and outside $G'$, $h$ agrees with $g$. Otherwise, we go on to the Second scenario. \\

\begin{figure}[H]
  \begin{center}
   \includegraphics[scale=0.55]{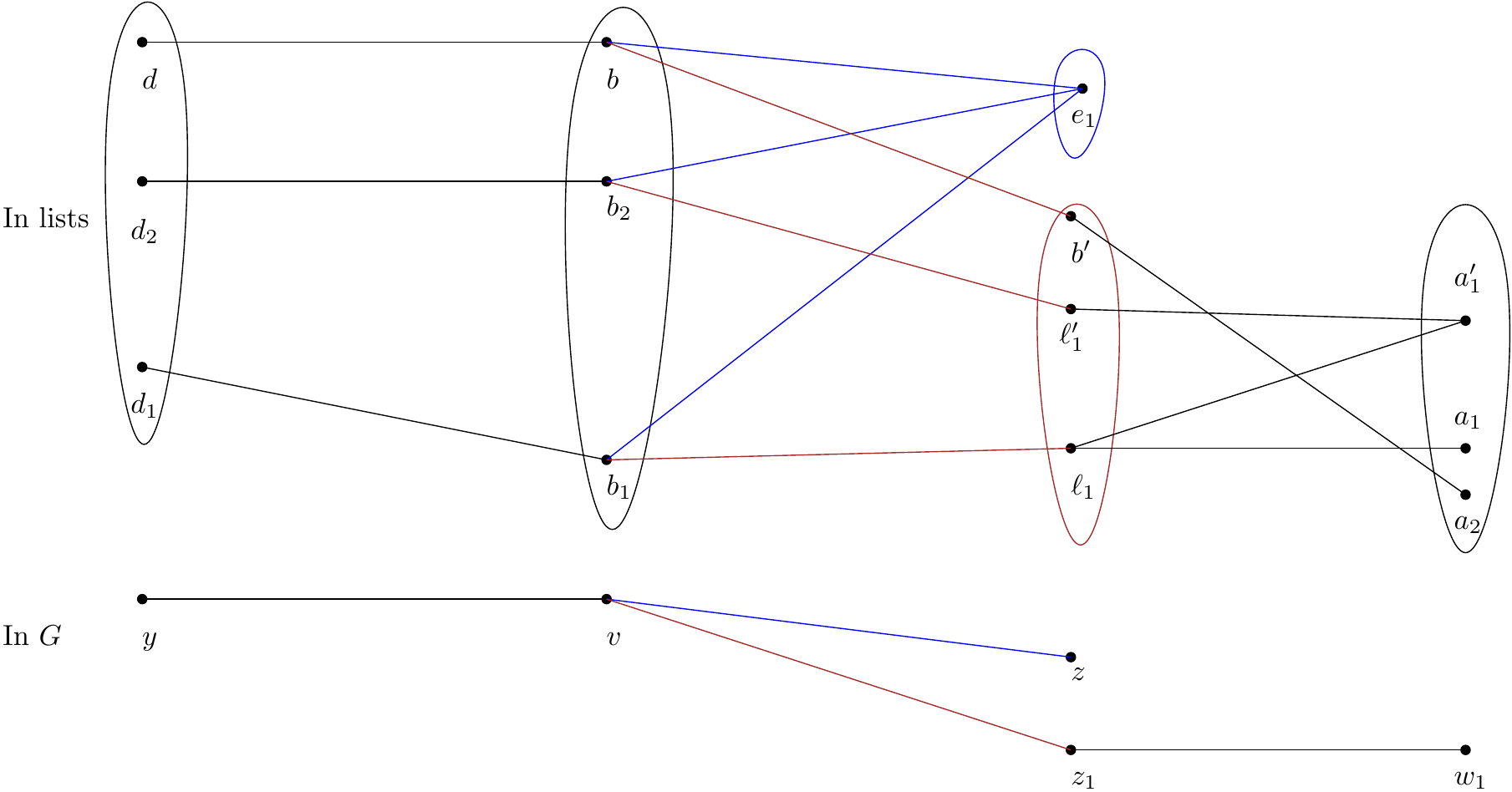}
  \end{center}
  \caption{ $z_1 \in B(G')$, and $\ell_1,\ell'_1 \in L_{z,e_1}(z_1)$. There is a walk from $b$ to $b'=f(z_1;(\ell'_1)^k,\ell_1)$}
\label{fig:Proof-fig4}
 \end{figure}

\noindent {\textbf{Second scenario.}} There is \textbf{no}  $\ell'_1 \in L_1(z_1) \cap L_{2}(z_1)$ 
such that $(b_1,\ell'_1),(b_2,\ell'_1) \in L(v,z_1)$ (see Figure \ref{fig:Proof-fig4}).  
At this point we need to start from vertex $v,b_1,b_2,b \in L(v)$. We consider the digraph $(G^2,L^2) =$ \Call{Sym-Diff}{$G,L_{2},v,b_1,b_2,z,e_1$} and follow homomrphism $g$ inside $G^2$, by considering  $B(G^2)$, and further modifying $g$ so that its images are reachable from $b$ simultaneously. For example, let $\ell_1=g(z_1)$, $\ell'_1 \in L(z_1)$ with $g_2(v_1)\ell'_1 \in A(H)$, where $v_1z_1$ is the last arc of $P_1$, and $b' \in L(z_1)$ with $b'=f(z_1;(\ell'_1)^k,\ell_1)$. Notice that here we may $g_2$ is an $L_2$-homomorphism obtained by calling \Call{Not-Minority}{$G_2,H,L_2,f$}. This homomorphism should exist because of Claim \ref{cl1} for $L_2$. 

Now as depicted in Figure \ref{fig:Proof-fig4}, let $w_1$ be a vertex in $B(G^2)$, and suppose there exists $a'_1 \in L_2(w_1) \cap L_1(w_1)$ such that $(\ell_1,a'_1) \in L_1(z_1,w_1)$ and $(\ell'_1,a'_1) \in L_2(z_1,w_1)$. 
Let $a_1 =g(w_1)$. Now as in Case 1,2, we conclude that 
there exists a path from $b'$ to $a_2= f(w_1;(a'_1)^k,a_1)$, and hence, we can further modify $g$ so that its image on $w_1$ is $a_2$. If there is no such $a'_1$ then we will be back in the Second scenario.

This process goes on as long as for all the boundary vertices the first scenario occur or we may reach to entire $G$. In any case, we would have a homomorphism that maps $y$ to $d$ and $z$ to $e_1$.

\begin{figure}[ht]
  \begin{center}
   \includegraphics[scale=0.6]{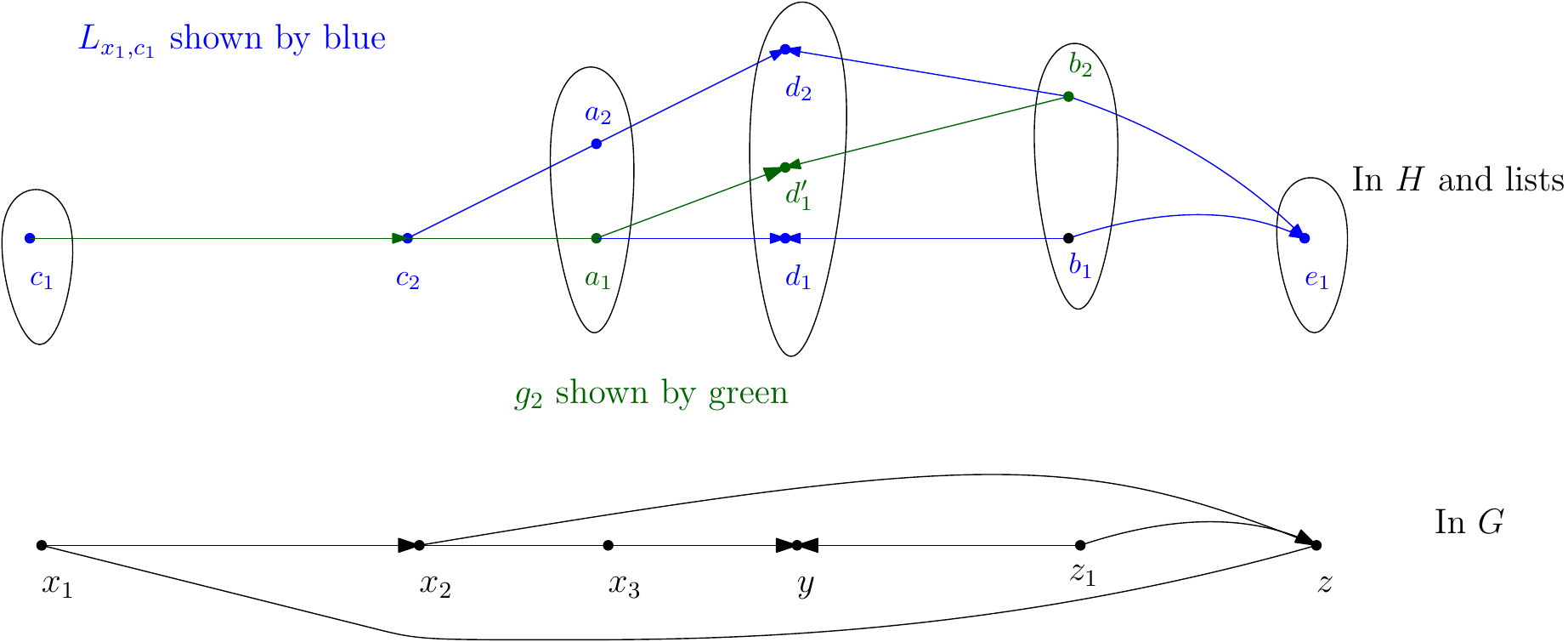}
  \end{center}
  \caption{ \small{Proof of Claim} \ref{cl1} }
\label{fig:small-test}
 \end{figure}

\begin{claim} \label{cl1}
Suppose all the small tests pass for instance $(G,L,H)$. Let  $x_1$ be an arbitrary vertex of $G$ and let $c_1 \in L(x_1)$. Let $L_1=L_{x_1,c_1}$. Then all the small tests pass for $G,L_1$. 
\end{claim}
\pf Let $G_1$ be the sub-digraph constructed in \Call{Sym-Diff}{$G,L,y,d_1,d_2,x_1,c_1$}. Note that there exists, an $L_1$-homomorphism, $g_1 : G_1 \rightarrow H$ with $g_1(x_1)=c_1$, $g_1(y)=d_1$. 
Let $L_2=(L_1)_{y,d_1}$, and  let $L_3=(L_2)_{x_2,c_2}$. 
The goal is to build a homomorphism $\psi$, piece by piece, that maps $x_1$ to $c_1$, $x_2$ to $c_2$, and $y$ to $d_1$ where its image lies in $L_3$. In order to to that we use  {\em  induction on the size of $\sum_{z_1 \in V(G_1)} |L_3(z_1)|$.} 
First suppose there exists $x_3$ on the oriented path from $x_2$ to $y$, and let $a_1,a_2 \in L_2(x_3)$ so that any oriented path from $a_1 \in L_1(x_3)$ inside $L_1(Y[x_3,x_2])$ ends at $c_2$. Since $d_1 \ne d_2$, it is easy to assume that $a_1 \ne a_2$. Now consider the sub-digraph, $G_2$ constructed in  \Call{Sym-Diff}{$G,L_1,x_3,a_1,a_2,x_1,c_1$} ( $a_1,a_2  \in L_2(x_3)$) and let $g_2$ be a homomorphism from $G_2$ to $H$. We may assume $g_2(y)=d'_1 \ne d_1$ (see Figure \ref{fig:small-test}). The other case; $d'_1=d_1$, is a special case of $d_1' \ne d_1$. 
Notice that by the choice of $x_3$, $g_2(x_2)=c_2$. Thus, $(c_2,d'_1) \in L_1(x_2,y)$. Let $L'_2=(L_1)_{x_2,c_2,x_3,a_1}$ and notice that $d_1 \in L'_2(y)$. The total size of all the lists of $L'_2$ is less than the total size of the lists in $L_1$ because $L'(x_3)=\{a_1\}$.  Thus, by induction hypothesis for the lists $L'_2$ we may assume that all the tests inside $L'_2$ pass, and hence, there exists an $L'_2$, homomorphism $g'_2$ from $G'_2$ to $H$, in which $g'_2(y)=d_1$. Here $G'_2$ is the sub-digraph constructed in \Call{Sym-Diff}{$G,L_1,y,d_1,d'_1,x_3,a_1$}. Notice that $x_3$ is in $B(G'_2)$. \\

\noindent {\textbf {Case 1.}} There exists a vertex of $z_1 \in B(G_2) \cap V(G'_2)$ (see Figure \ref{fig:small-test}). In this case, the homomorphism $\psi$ agrees with $g'_2$ on the path $Z$ from $y$ to $z \in B(G'_2)$ (where $z$ is also a vertex in $B(G_1)$) that goes through vertex $z_1$. \\

\begin{figure}[ht]
  \begin{center}
   \includegraphics[scale=0.6]{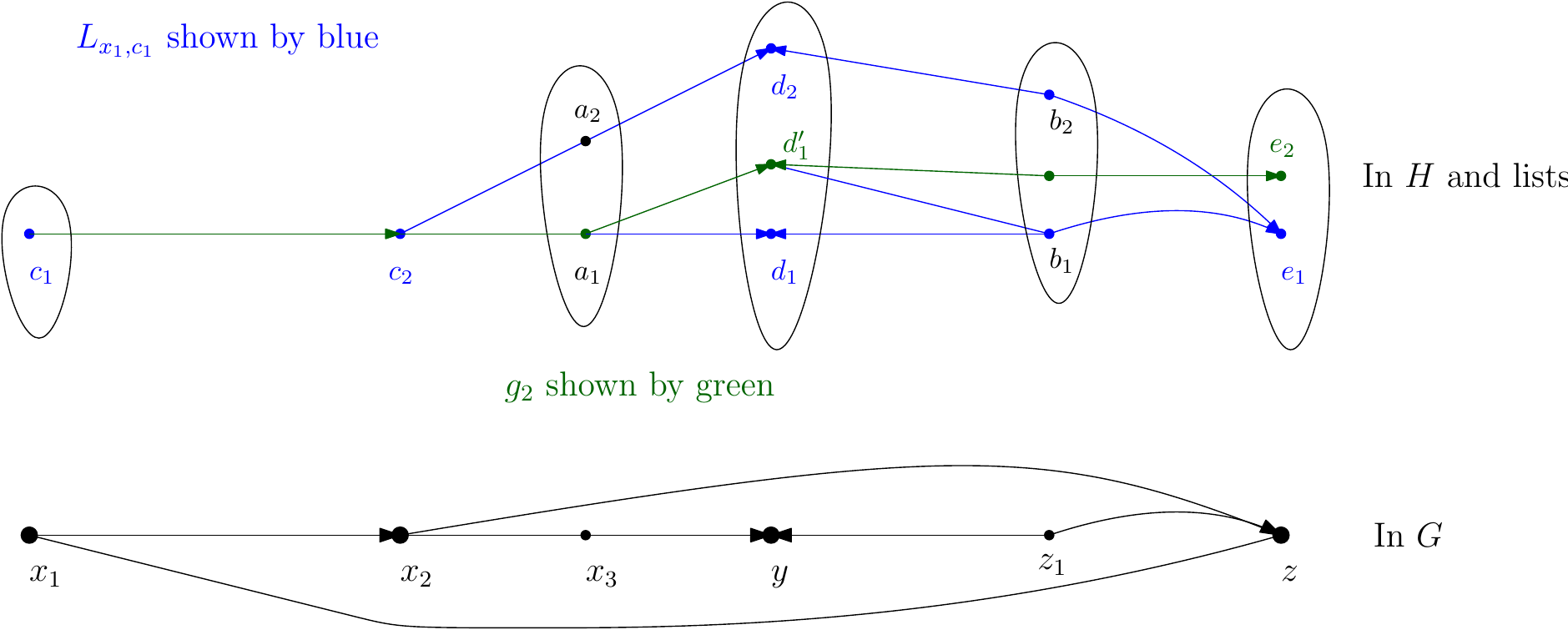}
  \end{center}
  \caption{ The homomorphism on $G'_2$ shown by green color  }
\label{fig:small-test2}
 \end{figure}


\noindent {\textbf {Case 2.}}  There exists a vertex of $z_1 \in B(G'_2)$, $z_1 \in V(G_2) \setminus B(G_2)$ (see Figure \ref{fig:small-test2}). Let $g'_2(z_1)=b_1$. Let $b_2 \in (L_1)_{x,c_2,y,d_2}(z_1)$. Now we consider the sub-digraph $G_3$ constructed in \Call{Sym-Diff}{$G,L_1,z_1,b_1,b_2,z,e_1$} where $z$ is a vertex in $B(G_1)$. There exists, a homomorphism $g_3$ from $G_3$ to $H$. We need to obtain $g'_3$ (using $g_3$ and induction hypothesis) so that its image on some part of $G_1 \cap (G_3 \cap G_2)$ lies inside $L_3$, and then $\psi$ would follow $g'_3$ on that part. To do so, we end with one of the cases 1, 2. If case 2 occurs we need to continue considering other partial homomorphisms. \\

Now consider the case where $x_3$ does not exist, in other words, $x_3$ is a vertex neighbor to $x_2$, and consider $a_1 \in L_3(x_3)$. In this case $c_2$ is adjacent to $a_1$, and we can extend the homomorphism $g_2$ to vertex $x_2$ where $g_2(x_2)=c_2$. 
\qed

\begin{lemma}
 The running time of the algorithm is $\mathcal{O}(|G|^4|H|^{k+4})$ (here $|G|$ is the number of arcs plus the number of vertices of $G$). 
\end{lemma}
\pf At the first glance, the algorithm \ref{alg-not-minority} 
looks exponential because we make polynomially many recursive calls. However, it is easy to see that the depth of the recursion is at most $|H|$. This is based according to the construction of  $(G',L')=$\Call{Sym-Diff}{$G,L,y,d_1,d_2,z,e_1$}; for each $v \in V(G')$, $|L'(v)| < |L(v)|$. This simply would guarantee that the running time of the algorithm is of $|G|^{\mathcal{O}(|H|)}$. 

However, it is more than just that, as the list becomes disjoint. Our implementation of algorithm would also support that the algorithm would perform much faster than $|G|^{\mathcal{O}(|H|)}$.  

According to Observation \ref{obs1}, 
we consider each connected component of $G \times_L H$ separately. The connected component partitioned the $G \times_L H$, and hence, the overall running time would be the sum of the running time of each connected components. We go through all the pairs and look for not-minority ones, which takes $\mathcal{O}(|G||H|^k)$ because we search for each $k$-tuple inside the list of each vertex $v$ of $G$. We also observe that the running time of the Preprocessing is $\mathcal{O}(|G|^3|H|^3)$. Note that at the end, we need to apply 
\Call{RemoveMinority}{} algorithm which we assume there exists one with running time $\mathcal{O}(|G|^3|H|^3)$ (see \cite{BD06}).

Now suppose Algorithm \ref{alg-not-minority} starts calling function \Call{Sym-Diff}{} where it 
considers two distinct vertices $y,z \in V(G)$ and $e_1 \in L(z)$, $d_1,d_2 \in L(y)$. The constructed instances are $T_1=(G_1,L_1)=$\Call{Sym-Diff}{$G,L,y,d_1,d_2,z,e_1$} and $T_2=(G_2,L_2)=$\Call{Sym-Diff} {$G,L,y,d_2,d_1,z,e_1}$. 
The lists $L_1,L_2$ (respectively) are disjoint because we exclude the boundary vertices into $G_1,G_2$. Therefore,
the running time of $T_1$ ($T_2$) is a polynomial of $\mathcal{O}(poly_1(|G_1|)*poly_2(|L_1|))$  ($ \mathcal{O}(poly_1(|G_2|)*poly_2(|L_2|)$) then the overall running time would be $\mathcal{O}(poly_1(|G|)*poly_2(|L|))$ for $G$ and $L$. Here $|L|$ denote the maximum size of a list and notice that $|L| \le |V(H)|$. 

Let $e_1,e_2,\dots,e_t \in L(z)$ and $d_1,d_2,\dots,d_r \in L(y)$ such that 
they induce a bi-clique in $L$. We may assume there exists at least one pair $d_1,d_2$ which is not minority. According to function \Call{Bi-Clique-Instances}{} instead of $(d_1,e_1)$ we use only $(d,e_1)$ where $d=f(y;d_2^k,d_1)$ and continue making the pair lists $L \times L$ smaller. Eventually each Bi-clique turns to a single path or the instance becomes Minority instance. Therefore, this step of the algorithm is a polynomial process with a overall running time  $\mathcal{O}(|G|^2|H|^{k+1})$ because we need to consider each pair of vertices of $G$ and find a bi-clique. This means the degree of the $poly_1$ is three and the degree of $poly_2$ is $k+1$ (we have of order $|G|^3$ from Preprocessing or from \Call{RemoveMinority}{} algorithm, and of order $|H|^{k+1}$ for finding bi-cliques).  
Therefore, the entire algorithm runs in $\mathcal{O}(|G|^4|H|^{k+4})$. This is because we consider every pair $x,y$, and $d_1,d_2 \in L(y)$, and $e_1 \in L(z)$. \qed

\section{ Weak near unanimity recognition algorithm} 

Let $A$ be a finite set. By a $k$-ary relation $R$ on set $A$ we mean a subset of the $k$-th cartesian power $A^k$; $k$ is said to be the \emph{arity}
of the relation. A constraint language (relational structure) $\Gamma$ over $A$ is a set of relations over $A$. A constraint language is finite if it contains finitely many relations. 

A hypergraph $\mathcal{G}$ on set $X$, consists of a set of hyperedges where each hyperedge $e$ is an {\em ordered} tuple $(x_1,x_2,\dots,x_k)$ , $x_1,x_2,\dots,x_k \in X$. Here $k$ is called the size of the hyperedge $e$. Notice that different hyperedges could have different sizes. 
A hypergraph is called uniform if all its hyperedges have the same size.  We denote the vertices of the hypergraph $\mathcal{G}$ by $V(\mathcal{G})$. 

For two hypergraphs $\mathcal{G},\mathcal{H}$, a homomorphism $f : \mathcal{G} \rightarrow \mathcal{H}$, is a mapping from $V(\mathcal{G})$ to $V(\mathcal{H})$ such that for every hyperedge $(x_1,x_2,\dots,x_k) \in \mathcal{G}$, $(f(x_1),f(x_2),\dots,f(x_k))$ is an hyperedge in  $\mathcal{H}$.

An instance of the constraint satisfaction problem (CSP) can be viewed as an instance of the hypergraph list homomorphism problem. We are given two hypergraphs $\mathcal{G},\mathcal{H}$ together with lists $\mathcal{L}$ where each hyperedge $\alpha \in \mathcal{G}$ has a list of possible hyperedges (all with the same size as $\alpha$) in $\mathcal{H}$, denoted by $\mathcal{L}(\alpha)$. 
The goal is to find a homomorphism  $f : \mathcal{G} \rightarrow \mathcal{H}$ such that for every hyperedge $\alpha=(x_1,x_2,\dots,x_k) \in \mathcal{G}$, $(f(x_1),f(x_2),\dots,f(x_k)) \in L(\alpha)$. In other words, if we look at the vertices of $\mathcal{G}$ as variables, the vertices of $\mathcal{H}$ as values, and hyperedges $\alpha's \in  \mathcal{G}$ as constraints then the existence of homomorphism $f$ illuminates a way of giving each variable a value, so that all constraints are satisfied simultaneously. A constraint is of form $(\alpha,L(\alpha))$, and a constraint is satisfied if tuple $\alpha$ is mapped by $f$ into one of the tuples in its list.

\begin{dfn} [Signature]
    For every two hyperedges $\alpha_1,\alpha_2$ from $\mathcal{G}$ ( or $\mathcal{H}$) we associate a signature 
    $S_{\alpha_1,\alpha_2}=\{(i,j) | \text{ $\alpha_1[i]=\alpha_2[j]$ } \}$ ( $\alpha_1[i]$ is the element in coordinate $i$-th of $\alpha_1$).  
  
\end{dfn}

Let $\mathcal{H}$ be a hypergraph on set $A$. Let $\mathcal{H}_1,\mathcal{H}_2,\dots,\mathcal{H}_t$ be a partition of $\mathcal{H}$ into $t$ uniform hypergraphs. 
A mapping $h : A^r \rightarrow A$ is a polymorphism of arity $r$ on $\mathcal{H}$ if $h$ is closed under each $\mathcal{H}_i$, $1 \le i \le t$. In other words, for every $r$ hyperedges $\tau_1,\tau_2,\dots,\tau_r \in \mathcal{H}_i$, $h(\tau_1,\tau_2,\dots,\tau_r) \in \mathcal{H}_i$. Notice that here $h$ is applied coordinate wise (e.g. if $(a_1,a_2,a_3),(b_1,b_2,b_3),(c_1,c_2,c_3)$ are hyperedges in $H_i$ then $(h(a_1,b_1,c_1),h(a_2,b_2,c_2),h(a_3,b_3,c_3))$ is a hyperedge in $H_i$). \\

In order to prove Theorem \ref{weak-nu-recognition} it is enough to prove the following theorem. 


\begin{theorem}\label{Maltsev-hypergrap} 
Let $\mathcal{H}$ be a hypergraph. Then the problem of deciding whether $\mathcal{H}$ admits a weak NU polymorphism is polynomial time solvable. 
\end{theorem}

The proof of the Theorem \ref{Maltsev-hypergrap} consists of several lemmas and an algorithm as follows. 
Let $\mathcal{H}_1,\dots,\mathcal{H}_t$ be a partitioned of $\mathcal{H}$ into uniform hypergraphs. We construct graph $G,H$ and lists $L$. \\

\noindent {\textbf {Vertices of $G,H$ and lists $L$:}} The vertices of $G$ are four tuples  $\overline{x}=(\alpha,\beta,\gamma,\lambda)$ where $\alpha,\beta,\gamma,\lambda \in \mathcal{H}_l$, $1 \le l \le t$. 
The vertices of $H$ are $\overline{\tau}$ where $\tau$ is a hyperedge of $\mathcal{H}$.  
For $\overline{x}=(\alpha,\beta,\gamma,\lambda)$, $L(\overline{x})$, consists of all $\overline{\tau}$, $\tau \in \mathcal{H}_l$ with the following property:

\begin{itemize}
    \item If $\alpha[i]=\beta[j]=\lambda[i]$, 
    $\alpha[j]=\gamma[j]=\beta[i]$, and $\gamma[i]=\lambda[j]$ then $\tau[i]=\tau[j]$. 
\end{itemize}

\noindent {\textbf {Adjacency in $G$ :}}
Two vertices $\overline{x}=(\alpha,\beta,\gamma,\lambda)$, and $\overline{y}=(\alpha',\beta',\gamma',\lambda')$ from $G$ with $\alpha,\beta,\gamma,\lambda \in \mathcal{H}_l$, $\alpha',\beta',\gamma',\lambda' \in \mathcal{H}_{l'} $ are adjacent if at least one of the following occurs. 

\begin{itemize}
    \item $S_{\alpha,\alpha'} \cap S_{\beta,\beta'} \cap S_{\gamma,\gamma'} \cap S_{\lambda,\lambda'} \ne\emptyset$. 
    
    \item $\alpha[i]=\beta'[j]=\lambda[i]$, 
    $\alpha'[j]=\gamma'[j]=\beta[i]$, and $\gamma[i]=\lambda'[j]$ 
\end{itemize}

\noindent {\textbf {Adjacency in $H$:}}
Two vertices $\overline{\tau} \in L(\overline{x})$ and $\overline{\omega} \in L(\overline{y})$ in $H$ where $\overline{x}=(\alpha,\beta,\gamma,\lambda)$, and $\overline{y}=(\alpha',\beta',\gamma',\lambda')$
are adjacent if the following occurs :

\begin{itemize}
    \item $S_{\alpha,\alpha'} \cap S_{\beta,\beta'} \cap S_{\gamma,\gamma'} \cap S_{\lambda,\lambda'} \subseteq S_{\tau,\omega}$
    
    \item If  $\alpha[i]=\beta'[j]=\lambda[i]$, 
    $\alpha'[j]=\gamma'[j]=\beta[i]$, and $\gamma[i]=\lambda'[j]$ then $\tau[i]=\omega[j]$. 
\end{itemize}

\begin{lemma}\label{sigger-G-H}
$\mathcal{H}$ admits a Siggers polymorphism if and only if there is an $L$-homomorphism from $G$ to $H$.
\end{lemma}
\pf Suppose $\mathcal{H}$ admits a Siggers polymorphism $\phi$. 
For every vertex $\overline{x}=(\alpha,\beta,\gamma,\lambda) \in G$ where $\alpha,\beta,\gamma,\lambda \in \mathcal{H}_l$, define mapping $g : G \rightarrow H$ with $g(\overline{x})=\overline{\phi(\alpha,\beta,\gamma,\lambda)}$ , where $\phi$ is applied coordinate wise. Let $\overline{y}=(\alpha',\beta',\gamma',\lambda')$ and suppose $\overline{x}\overline{y}$ is an edge of $G$. 
By definition, $\overline{\tau} \in L(\overline{x})$ where $\tau=\phi(\alpha,\beta,\gamma,\lambda)$ and 
$\overline{\omega} \in L(\overline{y})$ where 
$\omega =\phi(\alpha',\beta',\gamma',\lambda')$. Notice that if  $(i,j) \in S_{\alpha,\alpha'},S_{\beta,\beta'},S_{\gamma,\gamma'}$ then the value of $i$-th coordinate of $\tau$ is $\phi(a_1,a_2,a_3,a_4)$ ($a_1,a_2,a_3,a_4$ are the $i$-th coordinates of $\alpha,\beta,\gamma,\lambda$ respectively) and the value of the $j$-th coordinate of $\omega$ is $\phi(a_1,a_2,a_3,a_4)$ ($a_1,a_2,a_3,a_4$ are the $j$-th coordinate of $\alpha',\beta',\gamma',\lambda'$ respectively). Therefore, $(i,j) \in S_{\tau,\omega}$, and hence, there is an edge from $\overline{\tau}$ to $\overline{\omega}$ in $H$. Moreover, if  $\overline{x}$, $\overline{y}$ are adjacent because  $\alpha[i]=\beta'[j]=\lambda[i]$, 
    $\alpha'[j]=\gamma'[j]=\beta[i]$ then by definition $\tau[i]=\omega[j]$ and, and hence, $\overline{\tau}, \overline{\omega}$ are adjacent in $H$. 
Therefore, $g$ is a homomorphism from $G$ to $H$. 

Conversely, suppose $g$ is an $L$-homomorphism from $G$ to $H$. Suppose $\overline{\tau}=g(\overline{x})$ for $x=(\alpha,\beta,\gamma,\lambda)$. Then, 
for every $a_1,a_2,a_3,a_4$ that are the $i$-th coordinate of $\alpha,\beta,\gamma,\lambda$, respectively, set $\phi(a_1,a_2,a_3,a_4)=a_5$ where $a_5$ is the $i$-th coordinate of $\tau$ (recall $\tau$ is a ordered hyperedge corresponding to $\overline{\tau}=g(\overline{x})$). 

Consider a vertex $\overline{y} \in G$ with $\overline{y}=(\alpha',\beta',\gamma',\lambda')$. Suppose the $j$-coordinate of $\alpha',\beta',\gamma',\lambda'$ are $a_1,a_2,a_3,a_4$ respectively. Let $\overline{\omega}=g(\overline{y})$. By definition, $\phi(a_1,a_2,a_3,a_4)$ is $a'_5$ where $a'_5$ is the $j$-th coordinate of $\omega$. We show that $a_5=a'_5$. Observe that $(i,j) \in S_{\alpha,\alpha'} \cap S_{\beta,\beta'} \cap S_{\gamma,\gamma'} \cap S_{\lambda,\lambda'}$ where $a_1$ appears in $i$-th coordinate of $\alpha$ and in the $j$-th coordinate of $\alpha'$; $a_2$ is an element appearing in $i$-th coordinate of $\beta$ and in the $j$-th coordinate of $\beta'$; $a_3$ appears in the $i$-th coordinate of $\gamma$ and in the $j$-th coordinate of $\gamma'$; and finally $a_4$ appears in the $i$-th coordinate of $\lambda$ and in the $j$-th coordinate of $\lambda'$. 
Therefore, $\overline{x},\overline{y}$ are adjacent in $G$, and since $g$ is a homomorphism, $\overline{\tau}$ and $\overline{\omega}$ must be adjacent in $H$. By the construction of the lists, the $i$-th coordinate of $\tau$ is the same as the $j$-th coordinate of $\omega$, i.e. $a_5=a'_5$. 
Notice that since $g$ is a list homomorphism, 
$\overline{\tau} \in L(\overline{x})$ where $\tau= h(\alpha,\beta,\gamma,\lambda)$, and hence, $\tau$ belongs to $\mathcal{H}$. \qed \\

\begin{lemma}
If $\mathcal{H}$ admits a Siggers polymorphism then $G \times_L H^4$ admits a Siggers list polymorphism. 
\end{lemma}

\begin{lemma}
If $\mathcal{H}$ admits a Siggers polymorphism then it also admits a weak $k$-NU for some $k$ and $G \times_L H^k$ admits a weak $k$-NU list polymorphism.   
\end{lemma}
\pf It has been shown by Siggers \cite{Siggers} that if $\mathcal{H}$ admits a Sigger polymorphism it is also admit a weak NU of arity $k$ for some $k \ge 3$. It is easy to see that $G \times_L H^k$ admits a weak NU list polymorphis when $\mathcal{H}$ admits a weak $k$-NU.  \qed

Recall that for $a,b \in L(x)$, we say $(a,b)$ (with respect to $x$) is a minority pair if $f(x;a^k,b)=b$, otherwise, we call $(a,b)$ a non-minority pair. 

\begin{lemma}\label{lem:make-rectangle}
Let $x \in G$, $a,b \in L(x)$. Suppose there exists  $y \in G$, 
$c,d \in L(y)$ such that $(a,c),(a,d) \in L(x,y)$ and $(b,d) \in L(x,y)$ but $(b,c) \not\in L(x,y)$. Then at least one of the $(d,c)$, $(a,b)$ is not a minority pair. 
\end{lemma} 
\pf By contradiction suppose, $f(y;d^k,c)=c$. Now by applying the polymorphism $f$ on the $L(Y)$, where $Y$ is a path from $x$ to $y$, we conclude that $f(x;b,a^k) \ne b$ (because $(c,b) \not\in L(x,y)$), and hence, $f(x;a^k,b) \ne b$. \qed

\subsection{Algorithm for weak NU polymorphisms} 
\paragraph{Removing non-minority pairs}


Algorithm  \ref{alg-weak-nu} performs a test (with respect to $y,z \in G$ and $d_1,d_2 \in L(y)$, $e_1 \in L(z)$) on a sub-digraph $G'$ with the lists $L'$ which are constructed this way. Let $L_1=L_{z,e_1}$. 
First $G'$ includes vertices $v$ of $G$ such that for every $(d_1,j) \in L_1(y,v)$ we have $(d_2,j) \not\in L_1(y,v)$. 
Next, we further prune the lists $L'$ as follows; for each $v \in G'$, $L'(v)=\{i \mid (d_1,i) \in L_1(y,v)\}$. We ask whether there exists an $L'$-homomorphism from $G'$ to $H$. If there is no such homomorphism then we remove $(d_1,e_1)$ from the pair list $(y,z)$. 





  \begin{algorithm}
    \caption{Finding a homomorphism from $G$ to $H$}
  
   \label{alg-weak-nu}
    \begin{algorithmic}[1]
   
   \State \textbf{Input:} $G,H$, lists $L$ 
   
     \Function{Weak-NU}{$G,H,L$}
     
      \If { $\forall$ $x \in V(G)$, $|L(x)| \le 1$ }
         $\forall$ $x \in V(G)$ set $g(x)=L(x)$ 
       \State {\textbf {return}} $g$ \Comment{$g$ is a homomorphism from $G$ to $H$}
      \EndIf 
       
   \State If $G \times_L H$ is not connected then consider each connected component separately 
  
   
  
    
   
   \ForAll {$y,z \in V(G)$, $d_1,d_2 \in L(y)$, $e_1 \in L(z)$ s.t.
     $(d_1,e_1),(d_2,e_1) \in L(y,z)$ }
    \State $(G',L')=$\Call{Sym-Diff}{$G,L,y,d_1,d_2, z,e_1$} 
    \State $g^{y,z}_{d_1,e_1}=$ \Call{Weak-NU}{$G',H,L'$} 
    \If { $g^{y,z}_{d_1,e_1}$ is empty }  
          
          \State Remove $(d_1,e_1)$ from $L(y,z)$, and remove $(e_1,d_1)$ from $L(z,y)$ 
             
     \EndIf 
    
    \EndFor
    
     \State $g=$\Call{Getting-to-Minortiy}{$G,H,L$}
      
  
     

       \State {\textbf {return}} $g$  \Comment{$g$ is a homomorphism from $G$ to $H$}
    

   \EndFunction 
  \end{algorithmic}

 \end{algorithm}

\begin{algorithm}

 \begin{algorithmic}[1]
  \State \textbf{Input:} Digraphs $G$, lists $L$ and, $y,z \in V(G)$, $d_1,d_2 \in L(y)$, $e_1 \in L(z)$ 
 \Function{Sym-Diff} {$G,L,y,d_1,d_2, z,e_1$}
     
    %
     
     \State Create new lists $L'$ and let $L_1=L_{z,e_1}$, and construct the pair lists $L_1 \times L_1$ from $L_1$

      \State Set $L'(z)=e_1$, $L'(y)=d_1$, and set $G'=\emptyset$

      \ForAll { $v \in V(G)$ s.t. $\forall i$ with $(i,d_1) \in L_1(v,y)$ we have $(i,d_2) \not\in L_1(v,y)$ } 
      \State add $v'$ into set $G'$

     \EndFor  
     \State Let $G'$ be the induced sub-digraph of $G$

     
     
     
     
     
     \ForAll {$ u \in V(G')$} 
     
      \State $L'(u)=\{ \text {$i \mid   (d_1,i) \in L_1(y,u) $} \}$ 
     
     \EndFor
     \ForAll { $u,v \in V(G')$  }
       
\State $L'(u,v)=\{ (a,b) \in L_1(u,v) |  (a,b) \ \ is \ \ consistent \ \ in \ \ G' \} $.

     \EndFor

    \State {\textbf {return}} $(G',L')$

 \EndFunction

 \end{algorithmic}

 \end{algorithm}

\paragraph{Getting into minority pairs} 
For $x \in G$ and $a,b \in L(x)$, let $G^{L,x}_{a,b}$ be the set of vertices $y \in G$ such that for every $i \in L(y)$, $(a,i) \in L(x,y)$ and $(b,i) \not\in L(x,y)$. Let $B(G^{L,x}_{a,b})$ be the vertices of $G$ outside $G^{L,x}_{a,b}$ that are adjacent (via out-going, in-going arc) to a vertex in $G^x_{a,b}$.

\paragraph{Making Rectangles} Suppose for $a,b \in L(x)$, both $(a,b),(b,a)$ are minority pairs. We look at every vertex $y \in B(G^{L,x}_{a,b})$ and two vertices $c_1,c_2 \in L(y)$. Suppose $(a,c_1),(a,c_2),(b,c_2) \in L(x,y)$. Now by Lemma \ref{lem:make-rectangle} if $(b,c_1) \not\in L(x,y)$ then $(c_1,c_2)$ is not a minority pair and according to Algorithm \ref{alg-not-minority} (proof of correctness Lemma \ref{rec-apf-preserve} ($\lambda$)) we can remove, $(a,c_1)$ from $L(x,y)$. Therefore, as long as there exist such $y$ and $c_1,c_2 \in L(y)$ we continue doing so. At the end, we end up with an instance $G^{L,x}_{a,b} \cup B(G^{L,x}_{a,b})$ in which the rectangle property preserved for $x \in G$ and $a,b \in L(x)$ and any $y \in B(G^{L,x}_{a,b})$ and $c,d \in L(y)$. In other words, if $(a,c),(b,c),(a,d) \in L(x,y)$ then  $(b,d) \in L(x,y)$.  

\paragraph{Removing other non minority pairs} 
 Suppose for $a,b \in L(x)$, both $(a,b),(b,a)$ are minority pairs. Now for every $y,z \in B(G^{L,x}_{a,b})$ either $L_{x,a}(y,z) = L_{x,b}(y,z)$ or $L_{x,a}(y,z) \cap L_{x,b}(y,z)= \emptyset$. Otherwise, according to Lemma \ref{boundary}, we would have a non minority pair $(b_1,b_2)$ in $L(w)$, $w \in B(G^{L,x}_{a,b})$ and we can handle such non minority pairs according to function \Call{Clean-Up}{} in Algorithm \ref{alg-weak-nu}  (also according Algorithm \ref{alg-not-minority}). The proof of correctness of the function \Call{Clean-Up}{} appears in Lemma \ref{boundary} (3). Suppose we have removed non minority pairs $(b_1,b_2)$ with $b_1,b_2 \in L(w)$. Now we consider vertex $y \in G^{L,x}_{a,b}$, and two elements $c_1,c_2 \in L_{x,a}(y)$. Since, $(a,c_1),(a,c_2),(b,c_1),(b,c_2)$ are still in $L_{x,a}(x,y)$, $(c_1,c_2),(c_2,c_1)$ are still minority pairs, and hence we further consider $G^{L_{x,a},y}_{c_1,c_2}$, and $L_{x,a}$ and further remove other non minority pairs. We continue this process, until we reach the entire $G$, with the lists $L_{x,a}$. At this point ask the \Call{RemoveMinorty}{$G,H,L_{x,a}$}. If there is an answer then we have a homomorphism from $G$ to $H$ that maps $x$ to $a$ and we return such a homomorphism.

\begin{algorithm}

 \begin{algorithmic}[1]
  \State \textbf{Input:} Digraphs $G$, lists $L$ 
 \Function{Getting-to-Minority} {$G,H,L$ }
 \label{func:Handlin-Minority}     
    
 \State Let $x$ be a vertex in $G$ with $|L(x)| >1$

 \For { every two distinct $a, b \in L(x)$ }
 
 \State $L_1=L$ 
 
 
 
 \State $L'=$\Call{Clean-UP}{$G,H,L_1,x,a,b$}


\State $g=$ \Call{RemoveMinority}{$G,H,L'$}

\If {$g \ne \emptyset$ }
 \State {\textbf {return}} $g$
\EndIf 

\EndFor 

 \State  $g=$\Call{Getting-to-Minortiy}{$G,H,L_{x,a}$}
 \State {\textbf {return}} $g$

 
 


      

 \EndFunction

 \end{algorithmic}

 \end{algorithm}






\begin{algorithm}[ht]

 \begin{algorithmic}[1]
  \State \textbf{Input:} Digraphs $G$, lists $L$, $x \in V(G)$, $a,b \in L(x)$ 
 \Function{Make-Rectangle} {$G,H,L,x,a,b$ }
 \label{func:make-rectangle}     
    
 \For {$y \in G$, $c,d \in L(y)$}
 
 \If { $(a,c),(a,d),(b,d) \in L(x,y)$ and $(b,c) \not\in L(x,y)$}
 \State remove $(a,c)$ from $L(x,y)$
 \EndIf 
 
 \If { $(a,c),(b,c),(b,d) \in L(x,y)$ and $(b,d) \not\in L(x,y)$}
 \State remove $(a,c)$ from $L(x,y)$
 \EndIf

  \EndFor

 \State  \Call {PreProcessing}{$G,H,L$}

\State {\textbf {return}} $(L)$
    
 \EndFunction

 \end{algorithmic}

 \end{algorithm}

\begin{algorithm}[ht]

 \begin{algorithmic}[1]
  \State \textbf{Input:} Digraphs $G,H$, and lists $L$ 
 \Function{Clean-up} {$G,H,L,x',a',b'$ }
 \label{func:clean-up}     
 \State Let $St$ be an empty Stack. 
 
 \State push($x',a',b')$ into $St$.
 
 \While { $St$ is not empty}
  
  \State $(x,a,b)$ = pop($St$) and set
  Visit[$x,a,b$]=true
 
 \State \Call{Make-Rectangle}{$G,H,L,x,a,b$}
 \For {$y,z \in B(G^{L,x}_{a,b})$ }
 
 \State Let $c,d \in L(y)$ and $e,l \in L(z)$ s.t. $x,a,b$, $y,c,d$ and  $x,a,b$, $z,e,l$ are rectangles. 
 
 \State Let $v$ be a vertex at the intersection of $Y$ from $x$ to $Y$ and $Z$ from $x$ to $z$ in $G$.
 
 \State Let $a_1,a_2,a_3,a_4 \in L(v)$ such that  $(a,a_1),(a,a_3),(b,a_2),(b,a_4) \in L(x,v)$. 
  
 \If { $(a_1,e),(a_4,e),(a_2,l),(a_3,l) \in L(v,z)$ }
  \State Let $a_5 \in L(v)$ s.t. $(b,a_5) \in L(x,v)$, $(a_5,c) \in L_{x,b}(v,y)$ and $(a_5,e) \in L_{x,b}(y,z)$.

  \If {$\not\exists a_6 \in L(v)$ s.t. $(a,a_6) \in L(x,v)$, $(a_6,e) \in L_{x,a}(v,z)$, $(a_6,d) \not\in L_{x,a}(v,y)$  }
  \State remove $(a,d),(b,d)$ from $L(x,y)$
  \State \Call {PreProcessing}{$G,H,L$} 
  \EndIf
 
  \If { $\not\exists a_7 \in L(v)$ s.t. $(b,a_7) \in L(x,v)$, $(a_7,c) \in L_{x,b}(v,y)$, $(a_7,l) \not\in L_{x,b}(v,z)$  }
  \State remove $(a,l),(b,l)$ from $L(x,y)$ 
  
  \State \Call {PreProcessing}{$G,H,L$} 
  \EndIf
 
 \EndIf

 \If { $(a_1,e),(a_2,e),(a_3,l),(a_4,l) \in L(v,z)$ }
  \State Let $a_5 \in L(v)$ s.t. $(a,a_5) \in L(x,v)$, $(a_5,c) \in L_{x,a}(v,y)$, $(a_5,l) \in L_{x,a}(y,z)$.

  \If { $\not\exists a_6 \in L(v)$ s.t. $(a,a_6) \in L(x,v)$, $(a_6,l) \in L_{x,a}(v,z)$, $(a_6,d) \not\in L_{x,a}(v,y)$  }
  \State remove $(a,d),(b,d)$ from $L(x,y)$
  \State \Call {PreProcessing}{$G,H,L$} 
  \EndIf
 
  \If { $\not\exists a_7 \in L(v)$ s.t. $(a,a_7) \in L(x,v)$, $(a_7,c) \in L_{x,a}(v,z)$, $(a_7,l) \not\in L_{x,a}(v,y)$  }
  \State remove $(a,l),(b,l)$ from $L(x,y)$ 
  
  \State \Call {PreProcessing}{$G,H,L$} 
  \EndIf
 
 \EndIf 
 
  \EndFor 

 \If { $(a,d),(a,d),(b,c),(b,d) \in L(x,y)$ and Visit[$y,c,d]$=false }
  push($y,c,d$) into $St$ 
 \EndIf
 
  \If { $(a,e),(a,l),(b,e),(b,l) \in L(x,z)$ \hspace{1mm} and  Visit[$z,e,l]$=false }
  push($z,e,l$) into $St$ 
 \EndIf
 
\EndWhile 

\State {\textbf {return}} $(L)$
    
 \EndFunction

 \end{algorithmic}

 \end{algorithm}

\clearpage
\paragraph{Deploying Maltsev algorithm} 
Now we can deploy the procedure \Call{RemoveMinority}{} similar to Algorithm 1 in \cite{arash-maltsev}. The goal is to eliminate one of the $a,b$ from the list of $x$. We construct a sub-digraph of $G$ which is initially $G'=G^{L,x}_{a,b} \setminus B(G^{L,x}_{a,b})$ in function \Call{G-xab}{} (similar to Algorithm 1 in \cite{arash-maltsev}). If there are vertices $y,z \in B(G^{L,x}_{a,b})$ so that $L_{x,a}(y,z) \cap L_{x,b}(y,z) =\emptyset$, then we say $a,b$ are {\em not identical} on the boundary. In this case sub-digraph $G'$ will be extended from a vertex $y$ and two elements $c_1,c_2 \in L_{x,a}(y)$. This is done by adding $G^{L,y}_{c_1,c_2} \setminus B(G^{L,y}_{c_1,c_2})$ into $G'$ and the boundary vertices of $G'$ are updated by the vertices of  $B(G^{L,y}_{c_1,c_2})$ that are not already in $G'$.  
If the procedure \Call{RemoveMinority}{} finds a homomorphism that maps $x$ to $a$ in the sub-digraph $G'$ then we remove $b$ from $L(x)$, otherwise, we remove $a$ from $L(x)$. 
Following the proof of correctness of the Algorithm 1 in \cite{arash-maltsev} and Algorithm \ref{alg-not-minority} we obtain the correctness of this procedure. 

\paragraph{For every $a,b \in L(x)$, one of the $(a,b),(b,a)$ is not minority}
Let $a \in L(x)$ and let $L'_{x,a}$ be the subset of lists $L_{x,a}$ in which for every $y \in G$, and every $c_1,c_2 \in L(y)$, $(c_1,c_2)$ is a minority pair. We get to this point because there is no $L'_{x,a}$-homomorphism from $G$ to $H$ that maps $x$ to $a$. One reason is because at least one of the $(a,b),(b,a)$ is not a minority pair for $a,b \in L(x)$. If for every $c \in L(x)$, $(a,c)$ is not a minority pair then at the end we should see whether there exists an $L_{x,a}$-homomorphism from $G$ to $H$. 

\begin{algorithm}
  \caption{RemoveMinority -- Using Maltsev Property }
 
  \label{alg-remove-minority-new}
  \begin{algorithmic}[1]
  \Function{RemoveMinority}{$G,H,L$}

  \State \Call {Preprocessing}{$G,H,L$} and if a list becomes empty then {\textbf {return}} $\emptyset$ 
  
  \State Consider each connected component of $G \times_L H$ separately

  \Comment {we assume $G \times_L H$ is connected} 
  
  \State $\forall \ \ x \in V(G)$ and $\forall a,b \in L(x)$, if $a,b$ are twins then remove $b$ from $L(x)$.

 \ForAll{ $x \in V(G), a,b \in L(x)$ with $a \ne b$ }
 
   \State $(G',L')=$ \Call{G-xab}{$G,L,x,a,b$} \label{line5-m}
   
   \State $g^x_{a,b}=$ \Call{RemoveMinority}{$G',H,L'$} \label{line6-m}
    \If{  $g^x_{a,b}$ is empty } remove $a$ from $L(x)$ \label{line7-m}
    
    \Else {} remove $b$ from $L(x)$ \label{line8-m}
    
    \EndIf
    
    \State Preprocessing ($G,H,L$)
 \EndFor

\State Set $\psi$ to be an empty homomorphism.

 \If { $\exists x \in V(G)$ with $L(x)$ is empty }  return $\emptyset$. 
  \Else {} 
 \ForAll{ $x \in V(G)$ }  
 \State  $\psi(x)= L(x)$  \Comment{in this case the lists are singletons } 

 \EndFor

\EndIf

\State {\textbf {return} } $\psi$

\EndFunction

\end{algorithmic}
\end{algorithm}

\begin{algorithm}[ht]

 \begin{algorithmic}[1]
  \State \textbf{Input:} Digraphs $G$, lists $L$ and, $x \in V(G)$, $a,b \in L(x)$

\Function{G-xab} {$G,L,x,a,b$}

\State Set $\widehat{G^{L,x}_{a,b}}=G^{L,x}_{a,b} \setminus B(G^{L,x}_{a,b})$, and $B'=B(G^{L,x}_{a,b})$ 

\State Set stack $S$ to be empty  
and set $L_1= L_{x,a}$
\State push $x,a,b$ into $S$

\While{ $S$ is not empty }

\State pop $(x',a',b')$ from $S$

\ForAll{ $y \in B'$ and $c_1,c_2 \in L_1(y)$  s.t. $y,c_1,c_2$ witness $x',a',b'$ at $z,d_1,d_2$
}

\State Add new vertices from $G^{L_1,y}_{c_1,c_2} \setminus B(G^{L_1,y}_{c_1,c_2})$ into $\widehat{G^{L,x}_{a,b}}$ 

\State Update $B'$ by adding new boundary vertices from $B(G^{L_1,y}_{c_1,c_2})$ and removing the old 

\hspace{12mm} boundary  vertices from $B'$ that become internal vertices

\State push $(y,c_1,c_2)$ into $S$

\EndFor

\EndWhile

\State Initialize new lists $L'$
and  $\forall y \in \widehat{G^{L,x}_{a,b}}$, set $L'(y)=L_1(y)$


\State {\textbf{return}} $(\widehat{G^{L,x}_{a,b}},L')$ \Comment{ $L'$ lists should be $(2,3)$-consistent}

\EndFunction

 \end{algorithmic}

 \end{algorithm}

\begin{figure}[H]
  \begin{center}
   \includegraphics[scale=0.50]{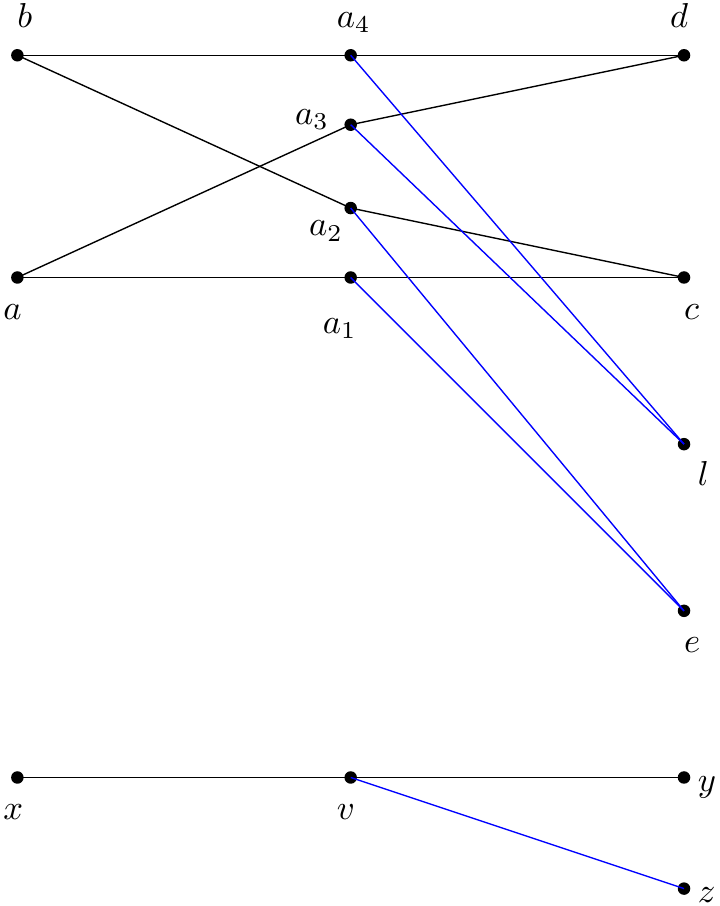}
  \end{center}
  \caption{\small{The pair lists on boundary vertices are the same.}}
  \label{fig:pair_list}
\end{figure}

\newpage
\begin{lemma}\label{boundary}
Suppose $a,b \in L(x)$ and both $(a,b)$, $(b,a)$ are minority pairs. 
Then one of the following occurs. 
\begin{enumerate}
    \item  For every $ y,z \in B(G^{L,x}_{a,b})$, 
    $L_{x,a}(y,z) =L_{x,b}(y,z)$, and $\forall w \in B(G^{L,x}_{a,b}), c_1,c_2 \in L_{x,a}(w)$,  $(c_1,c_2)$ is minority. 
    \item  For every $  y,z \in B(G^{L,x}_{a,b})$, 
    $L_{x,a}(y,z) \cap L_{x,b}(y,z) = \emptyset$, and $\forall w \in B(G^{L,x}_{a,b}), c_1,c_2 \in L_{x,a}(w)$,  $(c_1,c_2)$ is minority.   
    
    \item  There is $w \in B(G^{L,x}_{a,b})$ and  vertices $b_1,b_2 \in L_{x,a}(w)$ such that $(b_1,b_2)$ is not a minority pair. 
    
    
    \end{enumerate}
\end{lemma}
\pf  First assume for every $y \in B(G^{L,x}_{a,b})$ and every $c,d \in L_{x,a}(y)$, $(c,d)$ is a minority pair and every $c,d \in L_{x,b}(y)$, $(c,d)$ is a minority pair. 
Let $e,l \in L(z)$ such that $(a_1,e) \in L(v,z)$ and $(a_3,l) \in L(v,z)$. Notice that since $z \in B(G^{L,x}_{a,b})$, we must have a rectangle forming with $x,a,b$ and $z,c,d$. Without lose of generality we may assume that the internal vertices of this rectangle in $L(v)$ are $a_1,a_2,a_3,a_4$ (see Figure \ref{fig:pair_list}).

First suppose $(a_2,e),(a_4,l) \in L(v,z)$. 
Let $a'_1 \in L(v)$ such that $(a,a'_1) \in L(x,v)$ and $(a'_1,c) \in L(v,y)$ and suppose $(a'_1,t) \in L(v,z)$. Let $f(v;a'_1,a_1^{k-2},a_2)=a''_1$. 
By applying $f$ on $L(Y[v,x])[a'_1,a] ,\\ L(Y[v,x])[a_1,a], L(Y[v,x])[a_2,b], L(Z[v,z])[a_1,e], L(Z[v,z])[a_2,e], L(Z[v,z])[a'_1,t],$ we conclude that $(a''_1,b) \in L(v,x)$, $(a''_1,c) \in L(v,y)$, and $(a''_1,t) \in L(v,z)$. 
These mean that $(t,c) \in L_{x,a}(z,y)$ if and only if $(t,c) \in L_{x,b}(z,y)$. By symmetry, if there exists $a'_4 \in L(v)$ such that $(a'_4,d) \in L(v,y)$ and $(b,a'_4) \in L(x,y)$ then $(a''_4,t) \in L_{x,b}(z,y)$ if and only if $(a''_4,t) \in L_{x,a}(z,y)$, and hence, $(t,c) \in L_{x,a}(z,y)$ if and only if $(t,c) \in  L_{x,b}(z,y)$.
Now suppose $a'_2 \in L(v)$ such that $(b,a'_2) \in L(x,v)$, $(a'_2,c) \in L(v,y)$. Let $(a'_2,t) \in L(v,z)$ and let $f(v;a_1,a_2^{k-2},a'_2)=a''_2$. Again by applying the polymorphism $f$ on $L(Y[v,x])[a'_2,a] , L(Y[v,x])[a_2,a], L(Y[v,x])[a'_2,b]$ we conclude that $(a''_2,a) \in L(v,x)$. Therefore, $(t,c) \in L_{x,b}(y,z)$ if and only if $(t,c) \in L_{x,a}(y,z)$.

\begin{figure}[H]
  \begin{center}
   \includegraphics[scale=0.55]{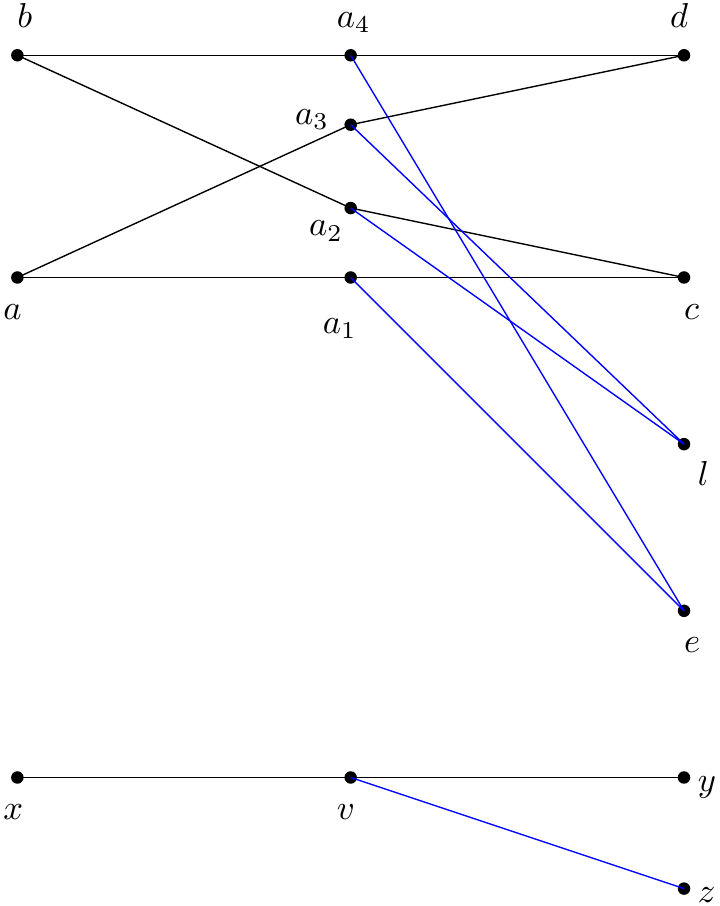}
  \end{center}
  \caption{ \small{The pair lists on boundary vertices are disjoint}}
  \label{fig:pair_list1}
\end{figure}

Second, suppose $(a_4,e),(a_2,l) \in L(v,z)$ (see Figure \ref{fig:pair_list1}).
Let $a'_1 \in L(v)$ such that $(a,a'_1) \in L(x,v)$ and $(a'_1,c) \in L(v,y)$ and suppose $(a'_1,t) \in L(v,z)$. Notice that $(t,c) \in L_{x,a}(z,y)$. Let $f(v;a'_1,a_1^{k-2},a_4)=a'_4$. By applying  $f$ on $L(Y[v,x])[a'_1,a], L(Y[v,x])[a_1,a], L(Y[v,x])[a_1,b],\\ L(Y[v,z])[a_1,e],L(Y[v,z])[a_4,e], L(Y[v,z])[a'_1,t],  L(Y[v,y])[a'_1,c], L(Y[v,y])[a_1,c],L(Y[v,x])[a_4,d]$,\\ we conclude that $(a'_4,b) \in L(v,x)$, $(a'_4,d) \in L(v,y)$,  $(a'_4,t) \in L(v,z)$, and hence, $(t,d) \in L_{x,b}(y,z)$. By symmetry, when $(b,a'_2) \in L(x,v)$, $(a'_2,c) \in L(v,y)$, and $(a'_2,r) \in L(v,z)$, then there exists $a'_3 \in L(v)$ such that $(a,a'_3) \in L(x,v)$, $(a'_3,b) \in L(v,y)$, and $(a'_3,r) \in L(v,z)$. We also observe that if $(a_3,e) \in L(v,z)$ then by rectangle property we have  $(a_1,l),(a_2,e),(a_4,l) \in L(v,z)$ and now it is easy to see that $L_{x,a}(y,z) =L_{x,b}(y,z)$ and (1) holds. Thus we assume that $(a_3,e) \not\in L(y,z)$ and it also follows that $(a_2,e),(a_1,l),(a_4,l) \not\in L(v,z)$. Since, the choice of $a_1,a_2,a_3,a_4$ are arbitrary, it is easy to see that $L_{x,a}(y,z) \cap L_{x,b}(y,z) = \emptyset$. \\

\begin{figure}[H]
  \begin{center}
   \includegraphics[scale=0.55]{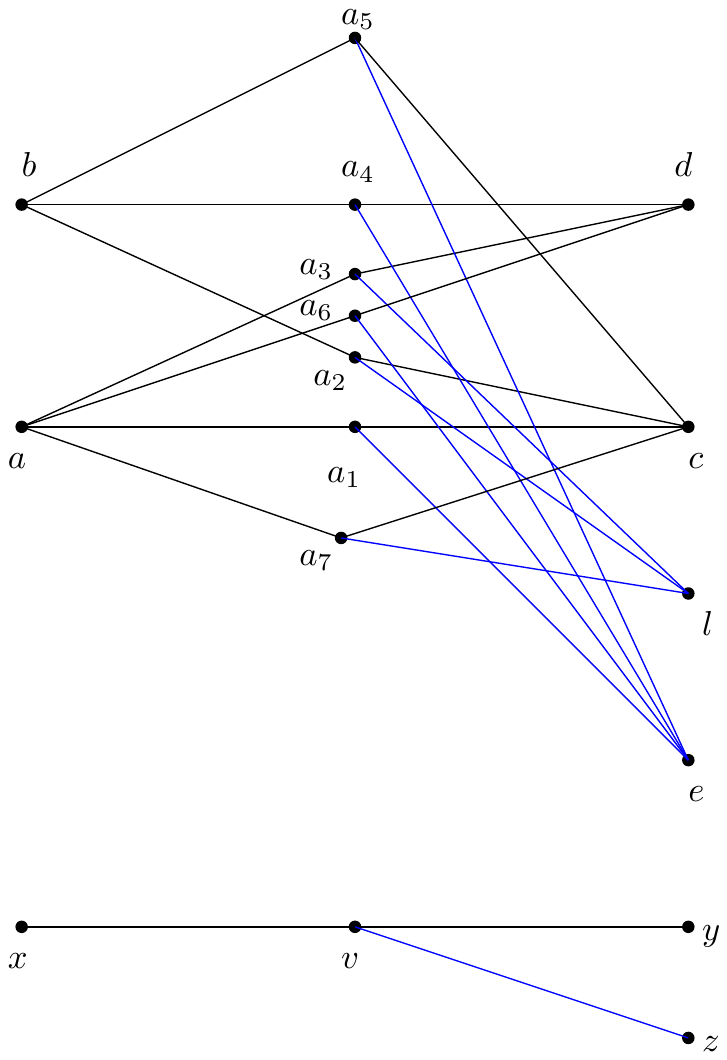}
  \end{center}
  \caption{ Finding a non minority pair .}
  \label{fig:attempt1}
\end{figure}

\noindent {\em Proof of 3}. Suppose none of the (2,3) occurs. Let $y,z \in B(G^{L,x}_{a,b})$, and let $Y$ be an oriented path from $x$ to $y$ in $G^{L,x}_{a,b}$ and $Z$ be an oriented path from $x$ to $z$. Let $v$ be a vertex in the intersection of $Z,Y$. Consider vertices $a_1,a_2,a_3,a_4 \in L(v)$ such that $(a,a_1),(a,a_3),(b,a_2),(b,a_4) \in L(x,v)$ and $(a_1,c),(a_2,c), (a_3,d),(a_4,d) \in L(v,y)$. 
First assume that there exist $e,l \in L(z)$ such that $(a_1,e),(a_4,e) \in L(v,z)$, and $(a_2,l),(a_3,l) \in L(v,z)$. In this case, we have $(e,c),(l,d) \in L_{x,a}(y,z)$, and $(e,d),(l,c) \in L_{x,b}(y,z)$. We may assume that $L_{x,a}(y,z),L_{x,b}(y,z)$ has an intersection. Without loss of generality, assume that there exists some $a_5 \in L_{x,b}(v)$ such that $(a_5,e) \in L_{x,b}(v,y)$,  $(a_5,c) \in L_{x,b}(v,y)$, and  $(e,c) \in L_{x,b}(z,y)$ (see Figure \ref{fig:attempt1}).

Let $a_6=f(v;a_1,a_5^{k-2},a_4)$. Notice that by applying $f$ on $Y[x,y]$, and $L(Y[x,y])$, 
starting at $(x;a^k,b)$ to $(v;a_1,a_5^{k-2},a_4)$, we conclude that there is a path in $L(Y[x,v])$ from $a$ to $a_6$. Moreover, by applying $f$ on $Y[v,z]$, and $L(Y[v,z])$, 
starting at $(x;a_1,a_5^{k-2},a_4)$ to $(z;e^k)$, we conclude that there is a path in $L(Y[v,z])$ from $a_6$ to $e$ so $(a_6,e) \in L_{x,a}(v,z)$. 
Finally, by applying the $f$ on $Y[v,y]$, $L(Y[v,y])$, we would get a path in $L(Y[v,y])$ from $a_6$ to $f(y;c^k,d)$. 
If $(a_6,d) \not\in L(v,y)$, then $(c,d)$ is not a minority pair, and hence, $w=z$, and $(b_1,b_2)=(c,d)$. Notice that if there exists at least one $a_6 \in L(v)$ so that $(a_6,e) \in L_{x,a}(v,z)$, $(a,a_6) \in L_{x,a}(x,v)$, $(a_6,d) \in L_{x,a}(v,y)$ then we can't conclude that $(c,d)$ is not a minority pair. So we may assume such $a_6$ exists, and we continue. Let $a_7=f(v;a_3,a_6^{k-2},a_1)$. Notice that since $f$ is a polymorphism, $(a,a_7) \in L_{x,a}(x,v)$. Now if $(a_7,c) \not\in L_{x,a}(v,y)$ then $f(x;d^k,c) \ne c$, and hence, $(d,c)$ is not a minority pair, and we set $w=y$, and $(b_1,b_2)=(d,c)$. Thus, we continue by assuming $(a_7,c) \in L_{x,a}(v,y)$. Now by following, $f$ on $Z[v,z]$, $L(Z[v,z])$ we conclude that $(a_7,f(v;e^k,l)) \in L_{x,a}(v,z)$. Therefore, if $(a_7,l) \not\in L_{x,a}(z,v)$ then $(e,l)$ is not a minority pair and we can set $w=z$, and $(b_1,b_2)=(e,l)$. Thus, we continue by assuming that $(a_7,l) \in L_{x,a}(v,z)$, and hence, $(l,c) \in L_{x,a}(z,y) \cap L_{x,b}(z,y)$.  

Next, let $f(v;a_2,a_5^k,a_4)=a_8$. Now by applying $f$, we conclude that $(b,a_8) \in L(x,v)$. By following, $f$ on $Z[v,z]$, $L(Z[v,z])$ we conclude that $(a_8,l) \in L_{x,b}(v,z)$, otherwise, $(e,l)$ is not minority pair and we are done. By 
following, $f$ on $Z[v,y]$, $L(Y[v,y])$, we conclude that $(a_8,d) \in L_{x,b}(v,y)$, otherwise, $(c,d)$ is not a minority pair. Therefore, $(l,d) \in L_{x,a}(z,y) \cap L_{x,b}(z,y)$. Since $a_5$ was an arbitrary vertex, we conclude that $L_{x,a}(y,z)=L_{x,b}(y,z)$, a contradiction to our assumption.   
The argument  when $(a_1,e),(a_2,e) \in L(v,z)$ and $(a_3,l),(a_4,l) \in L(v,z)$ is similar.  Note that when one of the (1,2) occurs on $B(G^{L,x}_{a,b})$ and (3) does happened and (3) does not happened then it is not difficult that we can assume that for every $y \in B(G^{L,x}_{a,b})$ and every $c_1,c_2 \in L_{x,a}(y)$, $(c_1,c_2)$ is a minority pair. 
\qed

\begin{lemma}
The Algorithm \ref{alg-weak-nu} correctly finds a homomorphism from $G$ to $H$ if one exists, and runs in polynomial time of $|G||H|$ and a polynomial of $|\mathcal{H}|$.
\end{lemma}
\pf The correctness of the algorithm follows from the correctness of the Algorithm \ref{alg-not-minority} and the algorithm in \cite{arash-maltsev}. Moreover, as we argue in the text before the algorithm, Lemma \ref{lem:make-rectangle}, and Lemma \ref{boundary}  justify the way we handle the minority pairs. The Algorithm in \cite{arash-maltsev} and Algorithm \ref{alg-not-minority} are polynomial of ( $poly(|G|)* poly(|H|)$).  
By the construction of $G,H,L$, the size of each list $L$ is a polynomial of $\mathcal{H}$. Therefore, it is not difficult to see that the Algorithm \ref{alg-weak-nu} runs in polynomial of $|\mathcal{H}|$. \qed

\section{Experiment}\label{experiment}
We have implemented our algorithm and have tested it on some inputs. The instances are mainly constructed according to the construction in subsection \ref{example}.

\begin{figure}[H]
  \begin{center}
   \includegraphics[scale=0.7]{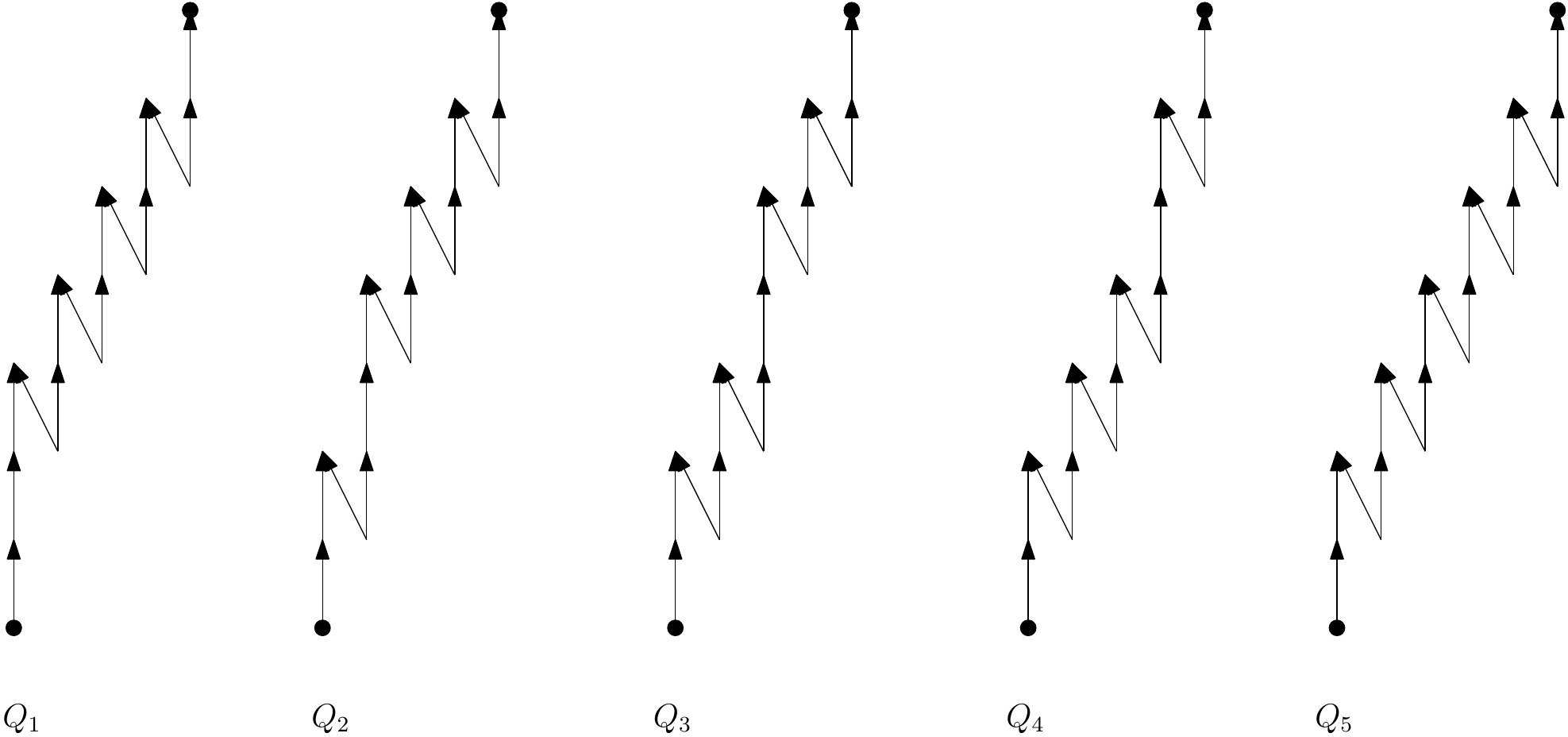}
  \end{center}
  \caption{Oriented paths of height 7}
\label{Q12345}
 \end{figure}

\begin{figure}
  \begin{center}
   \includegraphics[scale=0.62]{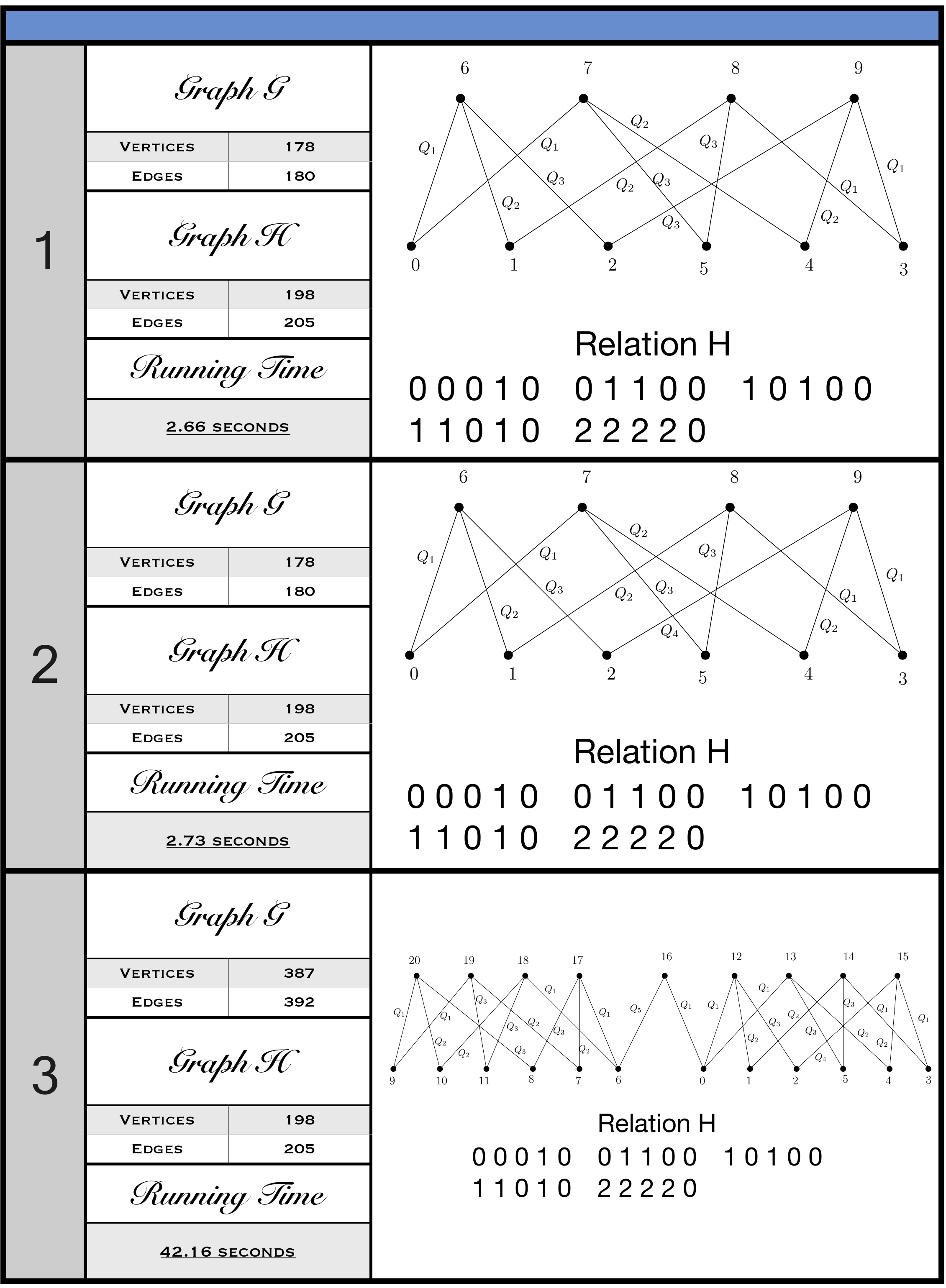}
  \end{center}
  \caption{$H$ is constructed from a relation of $5$-tuples and of arity $5$. $G$ is constructed from a bipartite graph where each edge in $G$ is replaced by one of the oriented paths depicted in Figure \ref{Q12345}.  }
\label{page1}
 \end{figure}

\begin{figure}
  \begin{center}
   \includegraphics[scale=0.62]{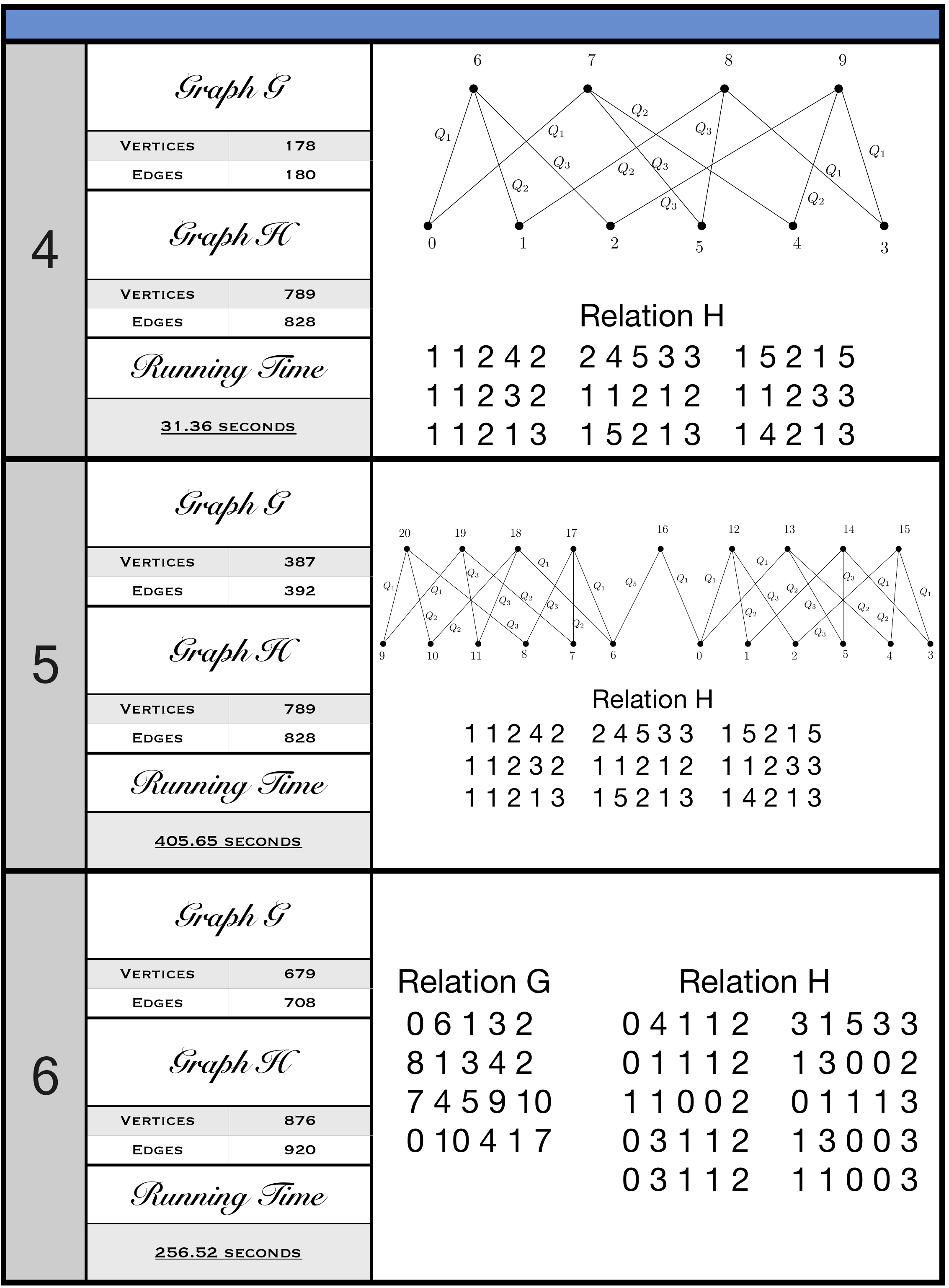}
  \end{center}
  \caption{First two $H$ digraphs constructed by a relation of $9$-tuples and of arity $5$ (which is closed under a semmi-lattice block Maltsev polymorphism, $01, 23, 45$). The first two $G$ digraphs 
  constructed from bipartite graphs by replacing their edges with oriented paths. The last instance, $G$ and $H$ are constructed from two relations according to the construction in Subsection \ref{example} }
\label{page2}
 \end{figure}
 
 \begin{figure}
  \begin{center}
   \includegraphics[scale=0.63]{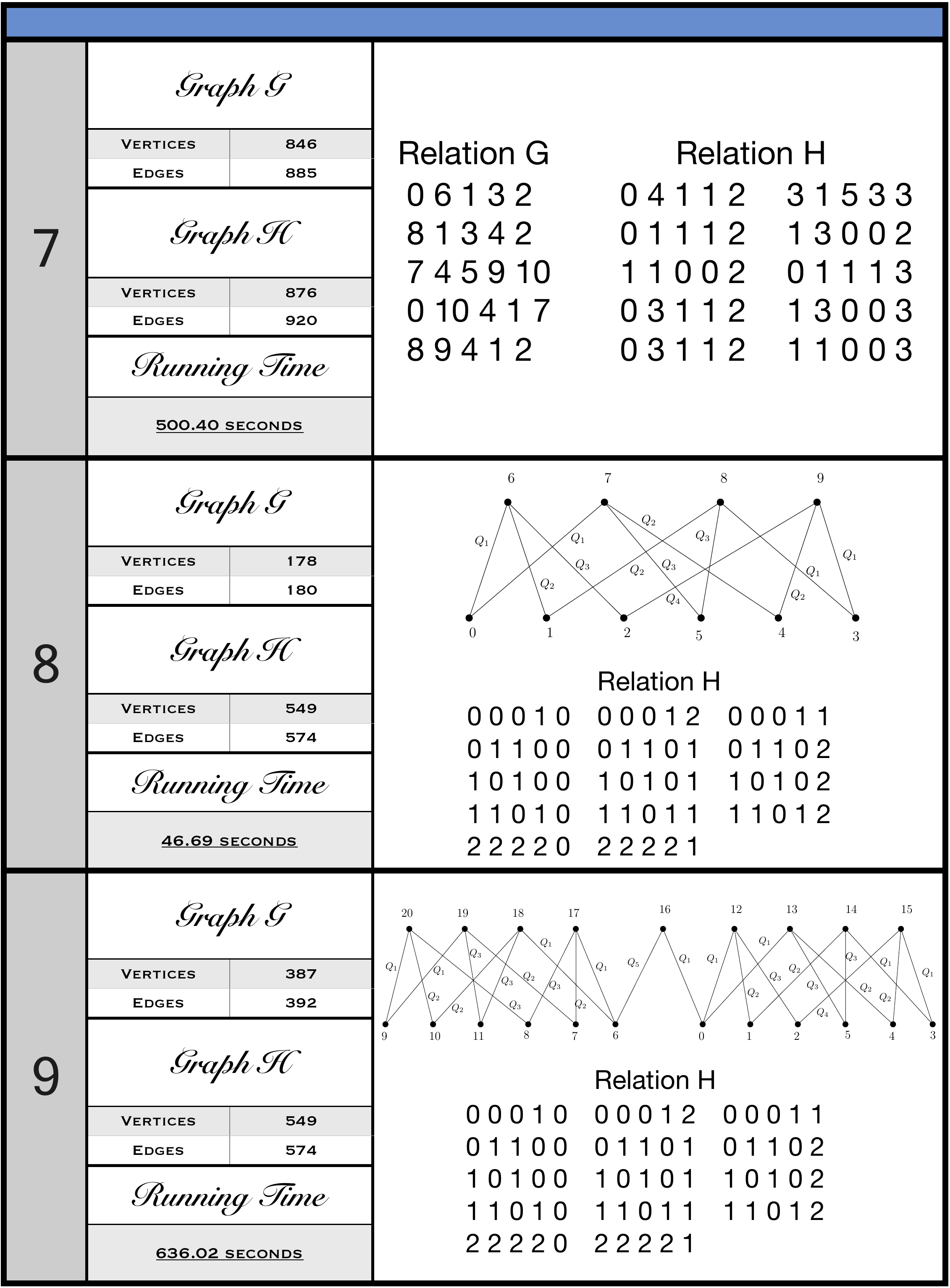}
  \end{center}
  \caption{ In the first instance, $G$ and $H$ are constructed from two relations according to the construction in Subsection \ref{example}.
  In the last two instances, $H$ is based on relation of $14$-tuples and arity $5$. $G$ is constructed from a bipartite graph using the oriented paths depicted in Figure \ref{Q12345}. }
\label{page3}
\end{figure}

 \begin{figure}
  \begin{center}
   \includegraphics[scale=0.63]{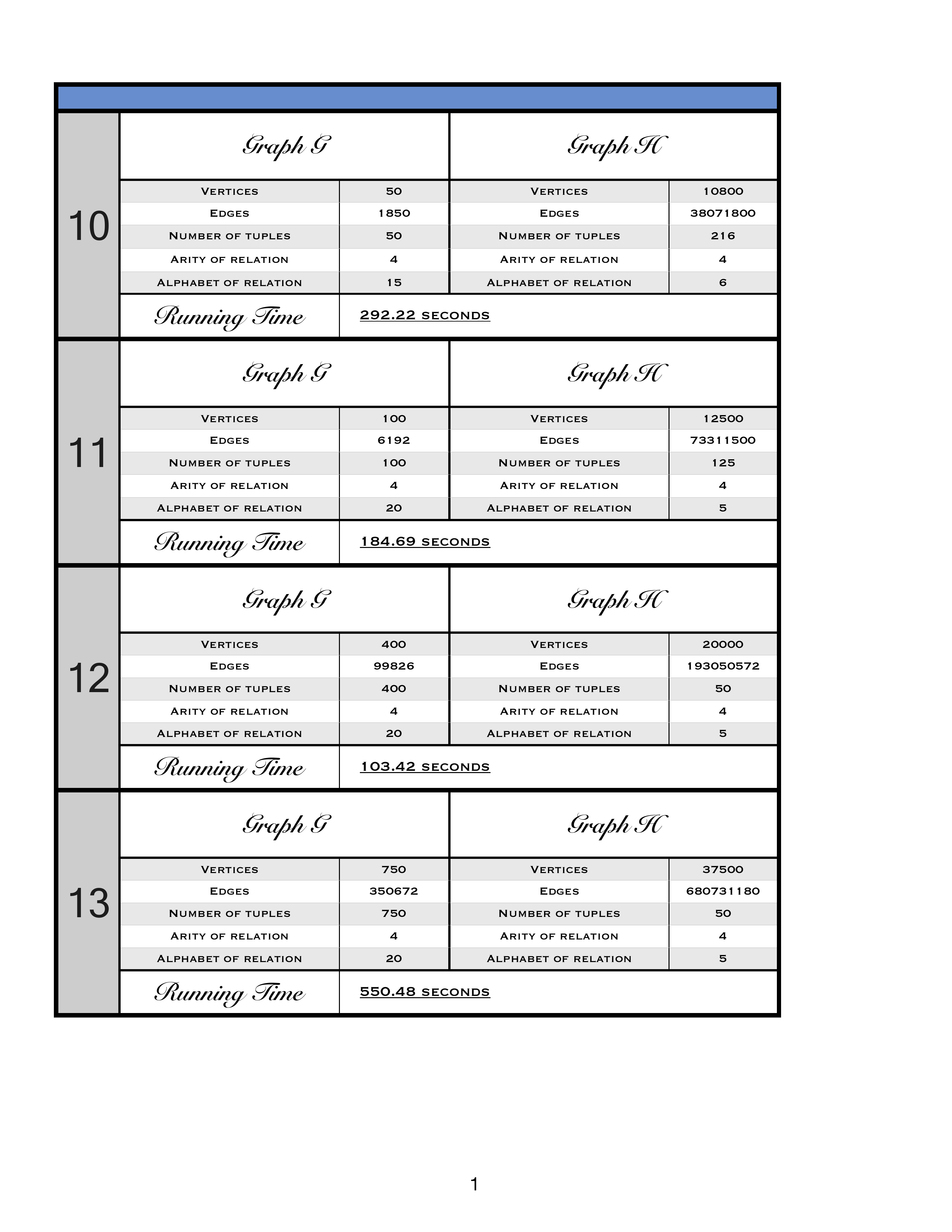}
  \end{center}
  \caption{ The examples constructed from  random relations $G$ and random target relations $H$.
  The $H$ relations are closed under a weak NU polymorphims of arity 3 which are not semilattice-block-Maltsev. 
   }
\label{page4}
\end{figure}

\newpage 
\subparagraph*{Acknowledgements:}
We would like to thank  V\'{i}ctor Dalmau, Ross Willard, Pavol Hell, and Akbar Rafiey for so many helpful discussions and useful comments.


{\small

\end{document}